\documentclass[10pt, conference, compsocconf]{IEEEtran}
\usepackage{algorithmicx}
\usepackage{algorithm}
\usepackage{tabularx}
\usepackage{colortbl}
\usepackage{algcompatible}
 \usepackage{algpseudocode}
\usepackage{graphicx}   \usepackage{tabularx}
\newcolumntype{s}{>{\raggedright\arraybackslash}X}
\usepackage{balance}    \usepackage{subfigure}
\usepackage{multirow}
\usepackage{amsmath}
\usepackage{cite}
\usepackage[table]{xcolor}\usepackage{epstopdf}
\usepackage{amsmath}
\usepackage{hhline}
\usepackage{setspace}
\usepackage[font={small}]{caption}
\newcommand{\system}{LaneQuest}
\def \sys {\textit{LaneQuest}}
\newcommand{\figscale}{0.7}
\definecolor{Gray}{gray}{0.9}
\definecolor{LightCyan}{rgb}{0.88,1,1}

\makeatletter

\IEEEoverridecommandlockouts
\begin{document}

\title{\system{}: An Accurate and Energy-Efficient Lane Detection System}

\author{
\IEEEauthorblockN{Heba Aly}
\IEEEauthorblockA{Computer and Sys. Eng. Dept.\\
Faculty of Eng., Alex. Univ., Egypt\\
Email: heba.aly@alexu.edu.eg}
\and
\IEEEauthorblockN{Anas Basalamah}
\IEEEauthorblockA{Comp. Eng. Dept.\& KACST GIS Tech. Innov. Ctr.\\
Umm Al-Qura Univ., KSA\\
Email: ambasalamah@uqu.edu.sa}
\and
\IEEEauthorblockN{Moustafa Youssef}
\IEEEauthorblockA{Wireless Research Center\\
E-JUST, Egypt\\
Email: moustafa.youssef@ejust.edu.eg}
\thanks{Moustafa Youssef is currently on sabbatical from Alexandria University, Egypt.}
}

\maketitle
\begin{abstract}

Current outdoor localization techniques fail to provide the required accuracy for estimating the car's lane.
In this paper, we present \sys{}: a system that leverages the ubiquitous and low-energy inertial sensors available in commodity smart-phones to provide an accurate estimate of the car's current lane. \sys{} leverages hints from the phone sensors about the surrounding environment to detect the car's lane. For example, a car making a right turn most probably will be in the right-most lane, a car passing by a pothole will be in a specific lane, and the car's angular velocity when driving through a curve reflects its lane. Our investigation shows that there are amble opportunities in the environment, i.e. lane ``anchors'', that provide cues about the car's lane. To handle the ambiguous location, sensors noise, and fuzzy lane anchors;  \sys{} employs a novel probabilistic lane estimation algorithm. Furthermore, it uses an unsupervised crowd-sourcing approach to learn the position and lane-span distribution of the different lane-level anchors.

Our evaluation results from implementation on different android devices and 260Km driving traces by 13 drivers in different cities shows that \sys{} can detect the different lane-level anchors with an average precision and recall of more than 90\%. This leads to an accurate detection of the exact car's lane position 80\% of the time, increasing to 89\% of the time to within one lane. This comes with a low-energy footprint, allowing \sys{} to be implemented on the energy-constrained mobile devices.
\end{abstract}

\section{Introduction}\label{sec:intro} 
\begin{figure}[!t]
\centering
\includegraphics[height=3cm,width=0.7\linewidth]{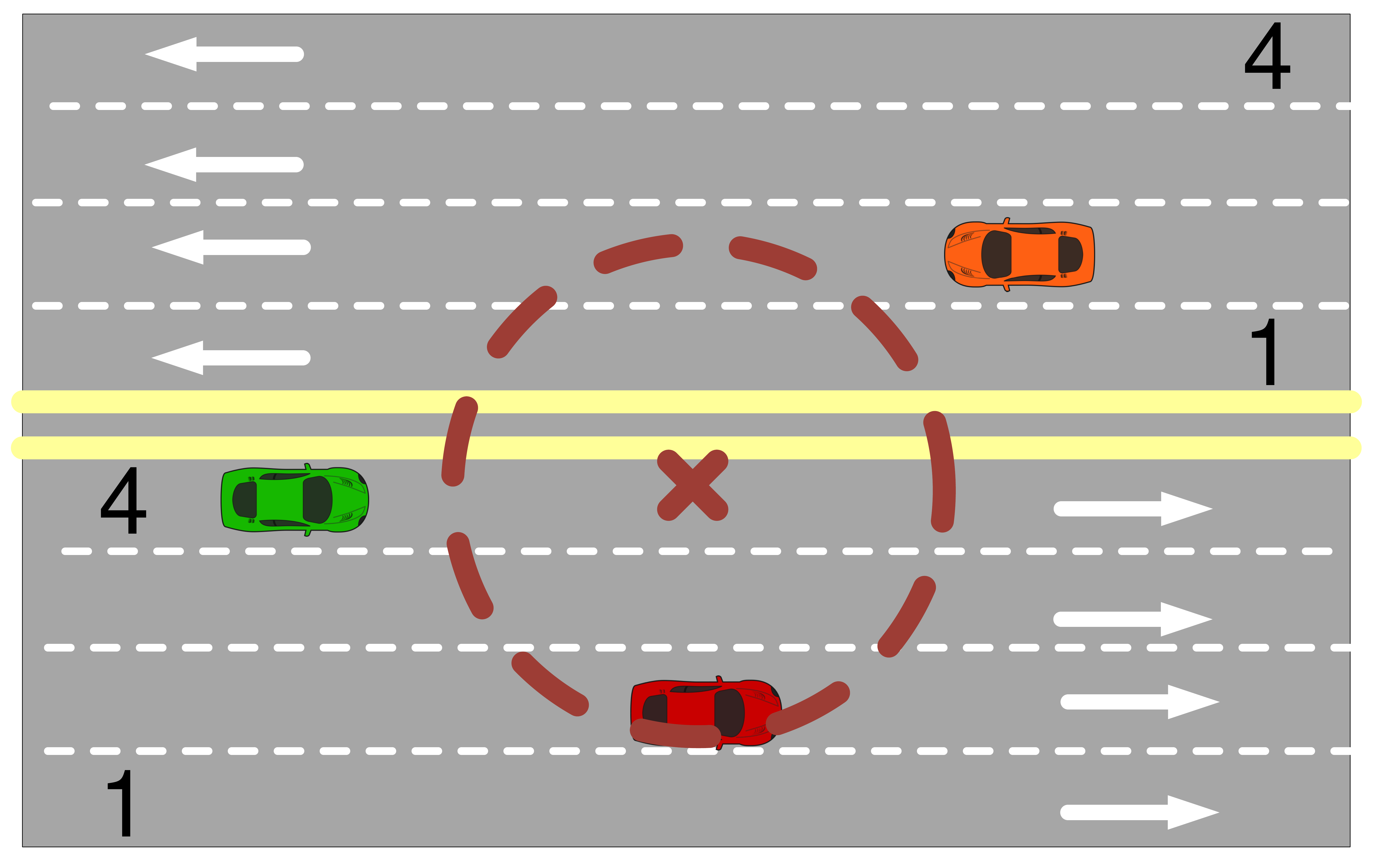}
\caption{Current outdoor localization technologies \emph{fail to provide enough accuracy to estimate the car's lane position}. The `$\times$' mark denotes the GPS position and the circle denotes the associated error. While the red car is moving in the 2nd lane, an error around three meters moves its estimate to the 4th lane.}
\label{fig:loc_error}
\end{figure}

Lane-level positioning systems for cars represent the next generation for outdoor navigation, where systems will not only predict the car location on the road but also its exact driving lane. This fine granularity is required for a wide range of emerging applications including advanced driver assistance systems (ADASs) \cite{hofmann2009quality}, autonomous cars (e.g. the Google driverless car~\cite{markoff2010google}), lane-level traffic estimation, electronic toll fee collection~\cite{de2011traffic}, predicting driver's intent~\cite{doshi2011tactical}, among others.  Current state-of-the-art outdoor car localization techniques can only provide an accuracy of about $10$ meters in urban environments~\cite{aly2013dejavu}. While such accuracy may be enough for ordinary location-based services \cite{ye2010location,aly_map14}, it fails to estimate the car's exact lane position (Figure~\ref{fig:loc_error}).

A number of systems were proposed to provide finer lane-level localization accuracy \cite{toledo2010lane,tao2013lane,ren2010lane}. However, these systems require special accurate GNSS devices (e.g. augmented GPS or RTK-GPS) and/or an expensive calibration phase (e.g. \cite{toledo2010lane,tao2013lane}), limiting their ubiquitous deployment. On the other hand, computer vision based techniques, e.g. \cite{ren2010lane}, use a camera to detect the lane markings. However, using an image processing solution raises accuracy challenges when road markings are unclear, line-of-sight is obstructed, and/or in bad weather conditions. It also requires extensive energy and processing power from commodity smart-phones.

In this paper, we present \sys{}; a system that leverages the ubiquitous sensors available in commodity smart-phones to provide an accurate and energy-efficient estimate of the car's current lane. Starting from an ambiguous location estimate, e.g. reported by the GPS, \sys{} leverages driving events detected by the phone sensors to reduce this ambiguity. Specifically, \sys{} uses the low-energy inertial sensors measurements to recognize unique motion events while driving such as changing the lane, turning right, or passing over a pothole. These events or ``lane anchors'' provide hints about the car current lane. For example, a car making a left turn most probably will be in the left-most lane; Similarly, potholes typically span only one lane, allowing detecting the lane of cars that pass through them. \sys{} uses a crowd-sensing approach to detect a large class of lane anchors as well as their positions through the road network and the lanes they span, exploiting them as opportunities for reducing the ambiguity in lane estimation.

To handle the sensors' noise, location ambiguity, and error in anchors location estimation, \sys{} models the lane estimation problem as a Markov lane detection problem that combines the car motion events (such as changing lanes) with lane anchor detection in a unified probabilistic framework. We have implemented \sys{} on different android devices and evaluated it using actual driving experiments at different cities using 13 drivers with a combined driving traces length of more than 260 km. Our results show that \sys{} can detect the different lane anchors with an average precision and recall of 95\% and 90\% respectively. This leads to accurately detecting the car lane more than 80\% of the time, increasing to 89\% to within one lane error. Moreover, \sys{} has a low-energy profile when implemented on top of different localization techniques.
In summary, our main contributions are four-folds:
\begin{itemize}
\item We present the architecture of \sys{}: an energy-efficient crowd-sensing system that leverages the sensed lane-anchors and car's dynamics to provide an accurate estimate of the car's current lane without any prior assumption on its starting lane position.
\vspace{-4pt}
\item We provide the details of a unified probabilistic framework for robust detection of cars' driving lane.
\vspace{-4pt}
\item We propose a crowd-sensing approach for detecting the position and lanes of different types of lane-level anchors. The proposed technique captures the inherent ambiguity in the crowd-sensing process.
\vspace{-4pt}
\item We implement \sys{} on android phones and evaluate its performance and energy-efficiency in different cities.
\end{itemize}

The rest of the paper is organized as follows: Section~\ref{sec:sys_overview} presents an overview of the system architecture. Section~\ref{sec:system} gives the details of the \sys{} system.  Section~\ref{sec:eval} provides our evaluation of \sys{}. We discuss the related work in Section~\ref{sec:relwork}. Finally, we conclude the paper and give directions for future work in Section~\ref{sec:conclude}.
\section{System Overview}\label{sec:sys_overview} \begin{figure}[!t]
\centering
\includegraphics[width=0.85\linewidth]{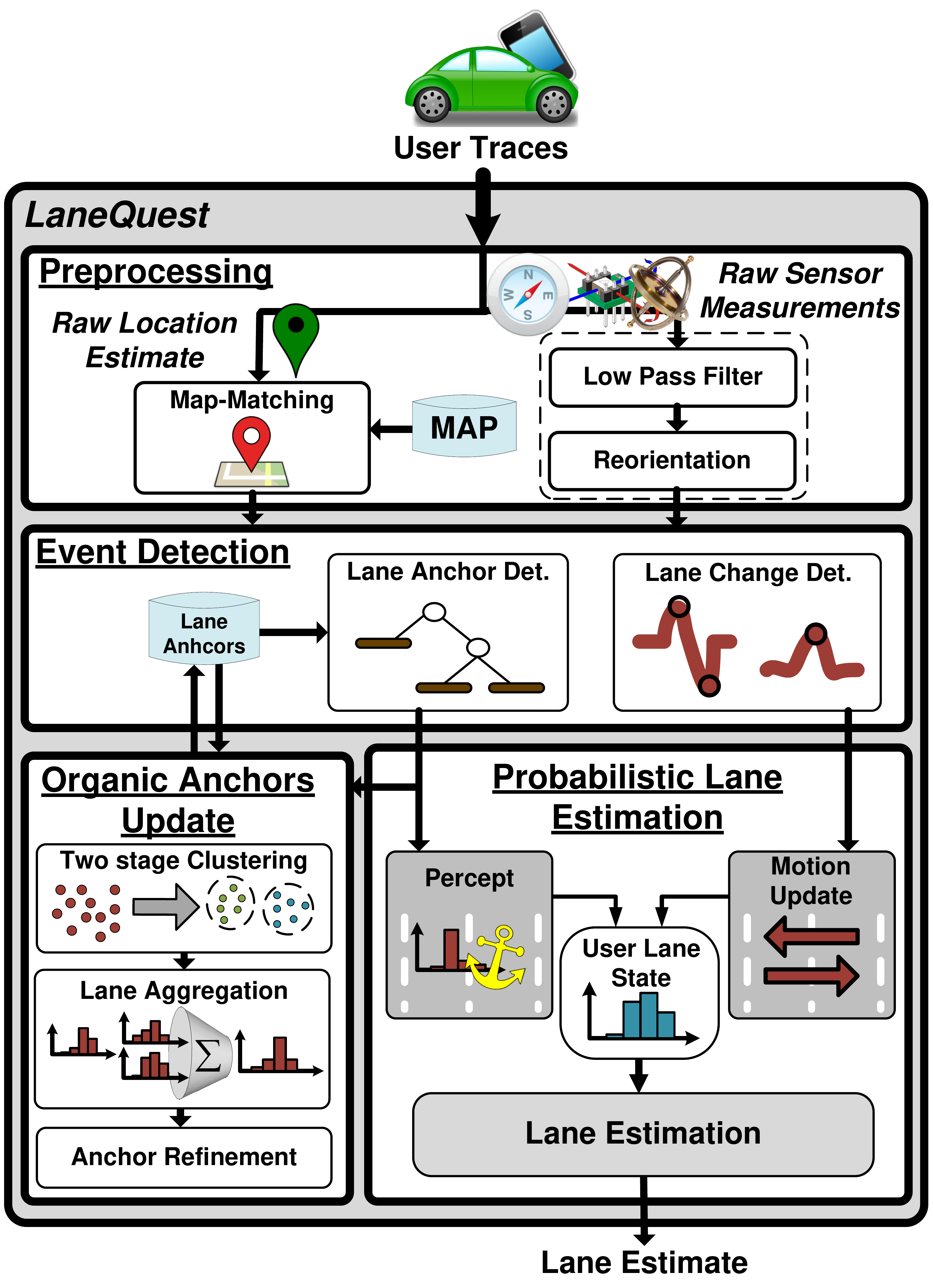}
\caption{\sys{} system architecture. \sys{} predicts the car current lane using a probabilistic approach by fusing knowledge of the car lane changes and a repository of lane-level anchors. Crowd-sourced traces are also used to detect new anchors and identify their lane position in an organic way.}
\label{fig:o_arch}
\end{figure}
Figure~\ref{fig:o_arch} shows an overview of the \sys{} system architecture. \sys{} estimates the car's lane position using inertial sensors available on a cell-phone attached to the car's windshield or a dashboard-mount. It leverages the car dynamics (e.g. changing lanes) and detected anchors in a probabilistic Markov framework to estimate the car's current lane. The system has four main components: the Preprocessing module, the Event Detection module, the Probabilistic Lane Estimation module, and the Lane Anchors Update module.
In this section, we give an overview of each of these modules.
\subsection{Preprocessing Module}

This module is responsible for preprocessing the raw input sensors and location data to reduce the noise effects. \sys{} collects time- and location- stamped measurements from the energy-efficient inertial sensors in the cell-phone. These include the accelerometer, gyroscope and magnetometer. To handle the noise in the sensors readings, we apply a local weighted low-pass regression filter~\cite{aly2013dejavu}. In addition, we also transform the sensor readings from the mobile coordinate system to the car coordinate system leveraging the inertial sensors~\cite{wang2013sensing,mohssen2014s}. After this transformation, the sensors y-axis points to the car direction of motion, x-axis to the left side of the car, and z-axis is perpendicular to Earth (pointing to the car ceiling).

For location information, \sys{} does not require a specific localization technique; it can leverage GPS, network-based localization techniques~\cite{cellsense,ibrahim2011hidden,ibrahim2010cellsense,ibrahim2013enabling,placelab} or other more accurate and energy-efficient GPS-replacement techniques, e.g. \cite{aly2013dejavu}. To further enhance the input location accuracy, we apply map matching~\cite{snapnet} to align the car's location estimates to the road network.
\begin{figure}[!t]
\centering
 \subfigure[Unknown Lane Position.]{
      \includegraphics[height=2.4cm,width=0.29\linewidth]{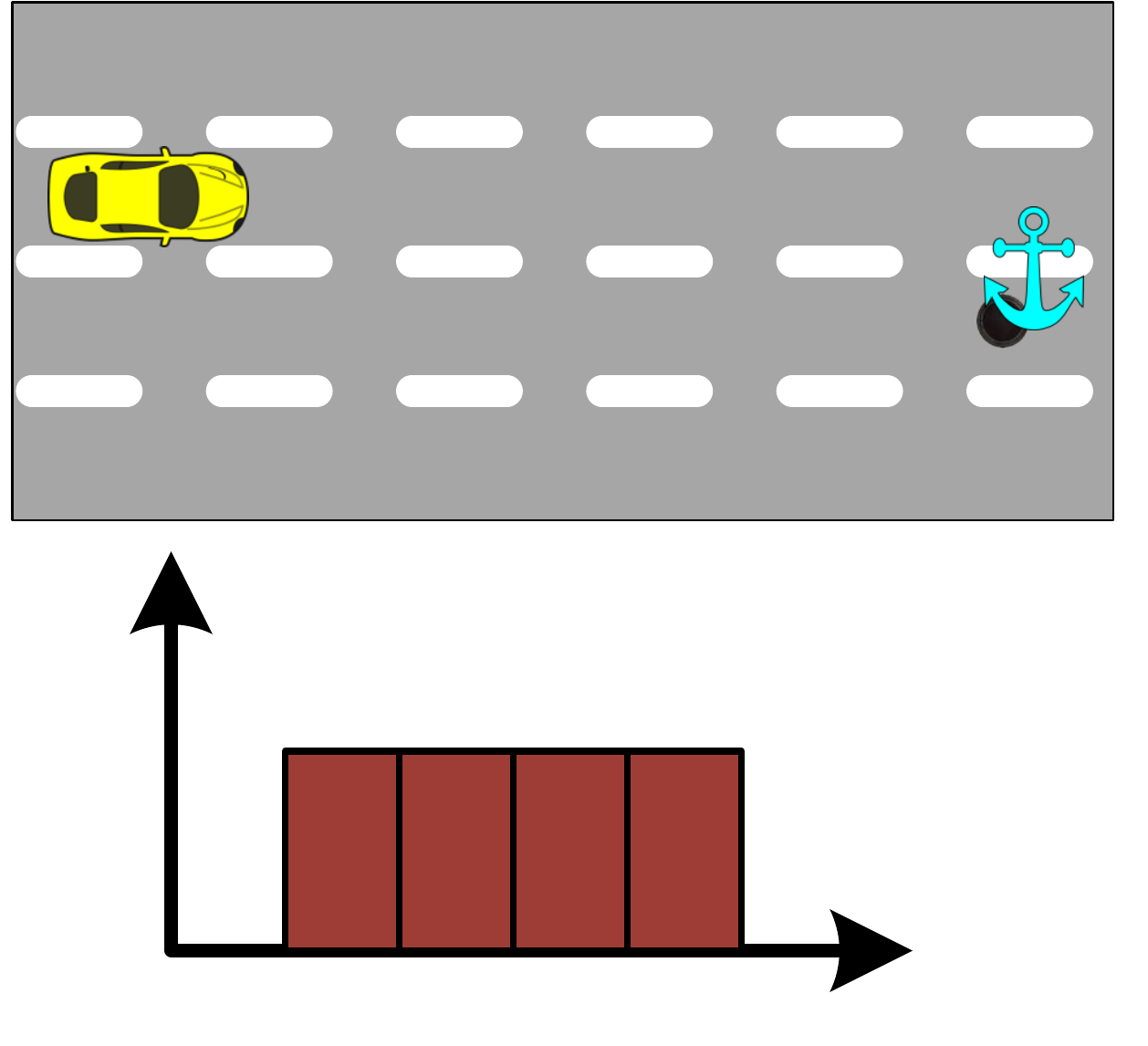}
      \label{fig:markov_ex_1}
    }
    \subfigure[Most probably not at the fourth lane.]{
      \includegraphics[height=2.4cm,width=0.29\linewidth]{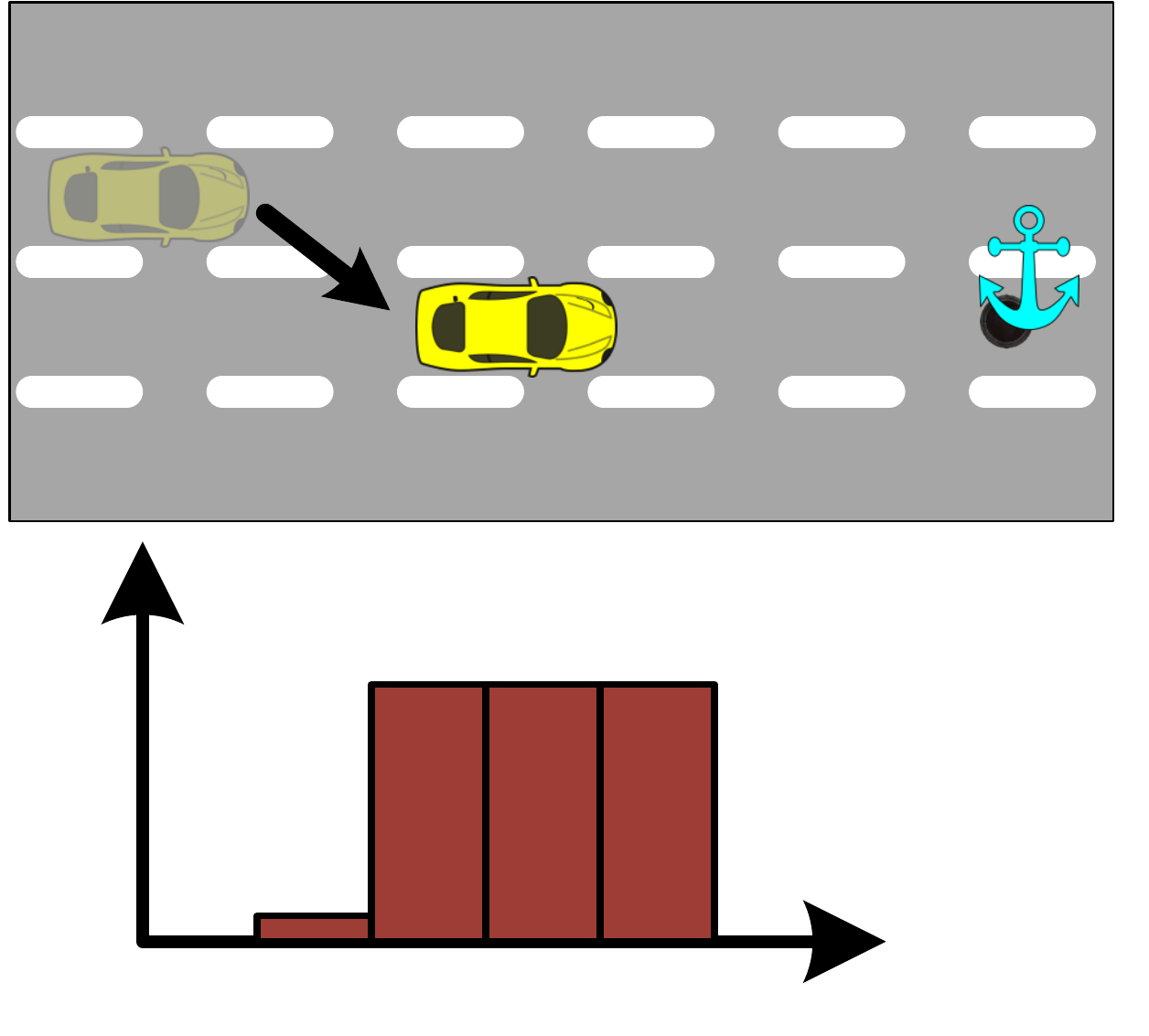}
      \label{fig:markov_ex_2}
    }
    \subfigure[Most probably at the second lane.]{
      \includegraphics[height=2.4cm,width=0.29\linewidth]{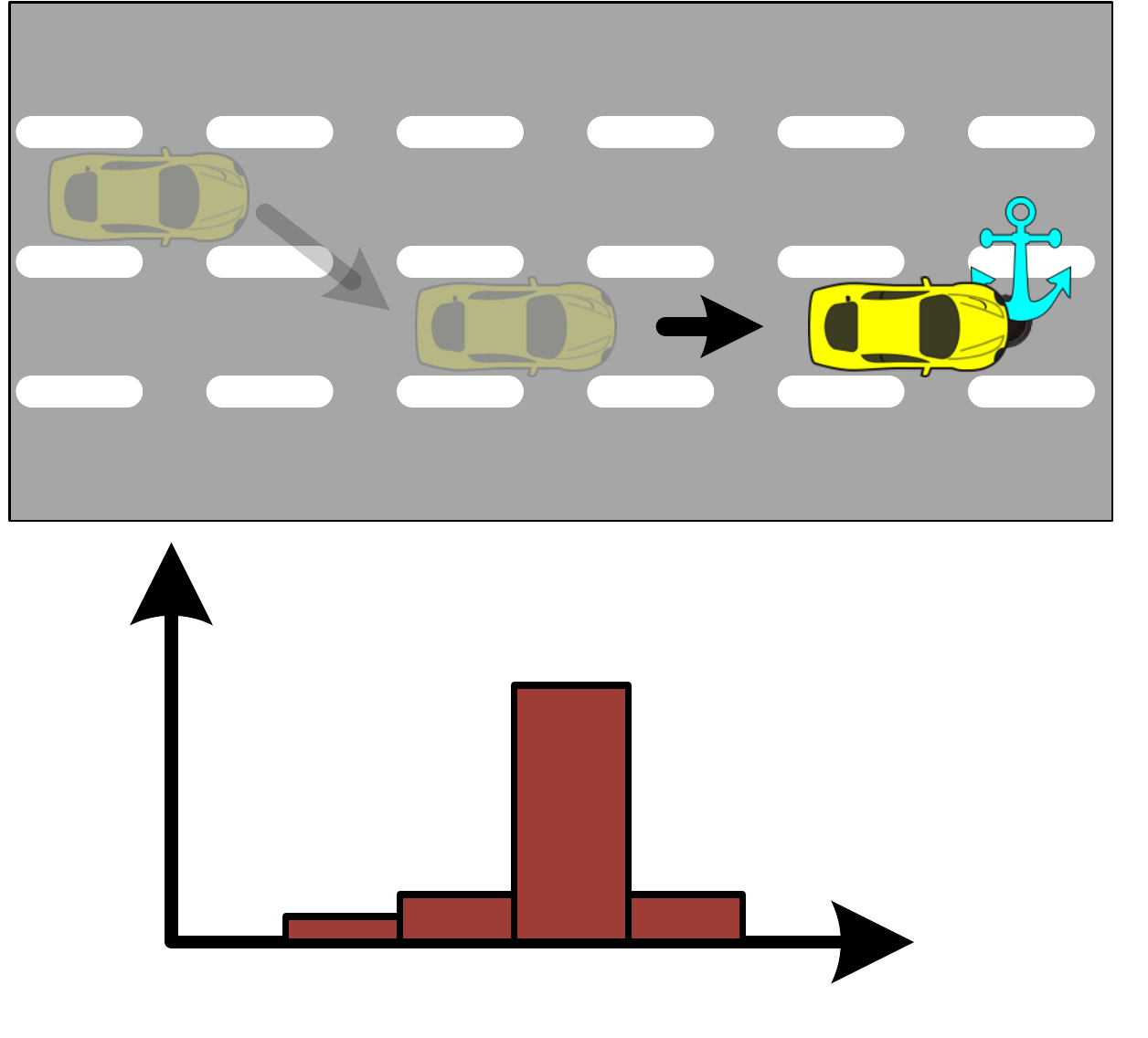}
  \label{fig:markov_ex_3}
    }
\caption{The probabilistic lane estimation basic idea: At the beginning, the car's lane is unknown. Then, as the car moves to the adjacent right lane the distribution shifts to the right since, most probably, the car is not at the left-most lane anymore. Finally, as the car encounters a lane anchor, its lane is mostly known as the anchor's lane.}
\label{fig:markov_ex}
\end{figure}
\subsection{Event Detection Module}
There are many driving patterns that can give cues for the car's current lane based on their unique signature on the different phone sensors. For example, when a car moves to the adjacent lane to the right, the car is with high probability not in the left-most lane (Figure~\ref{fig:markov_ex}). This ``lane change event'' can be detected by the phone inertial sensors (using the \emph{Lane Change Detection} sub-module) and used to reduce the ambiguity in the car's current lane.

\begin{figure}[!t]
\centering
 \subfigure[A car doing a u-turn will be at the left-most lane with high probability.]{
      \includegraphics[height=2.4cm,width=0.4\linewidth]{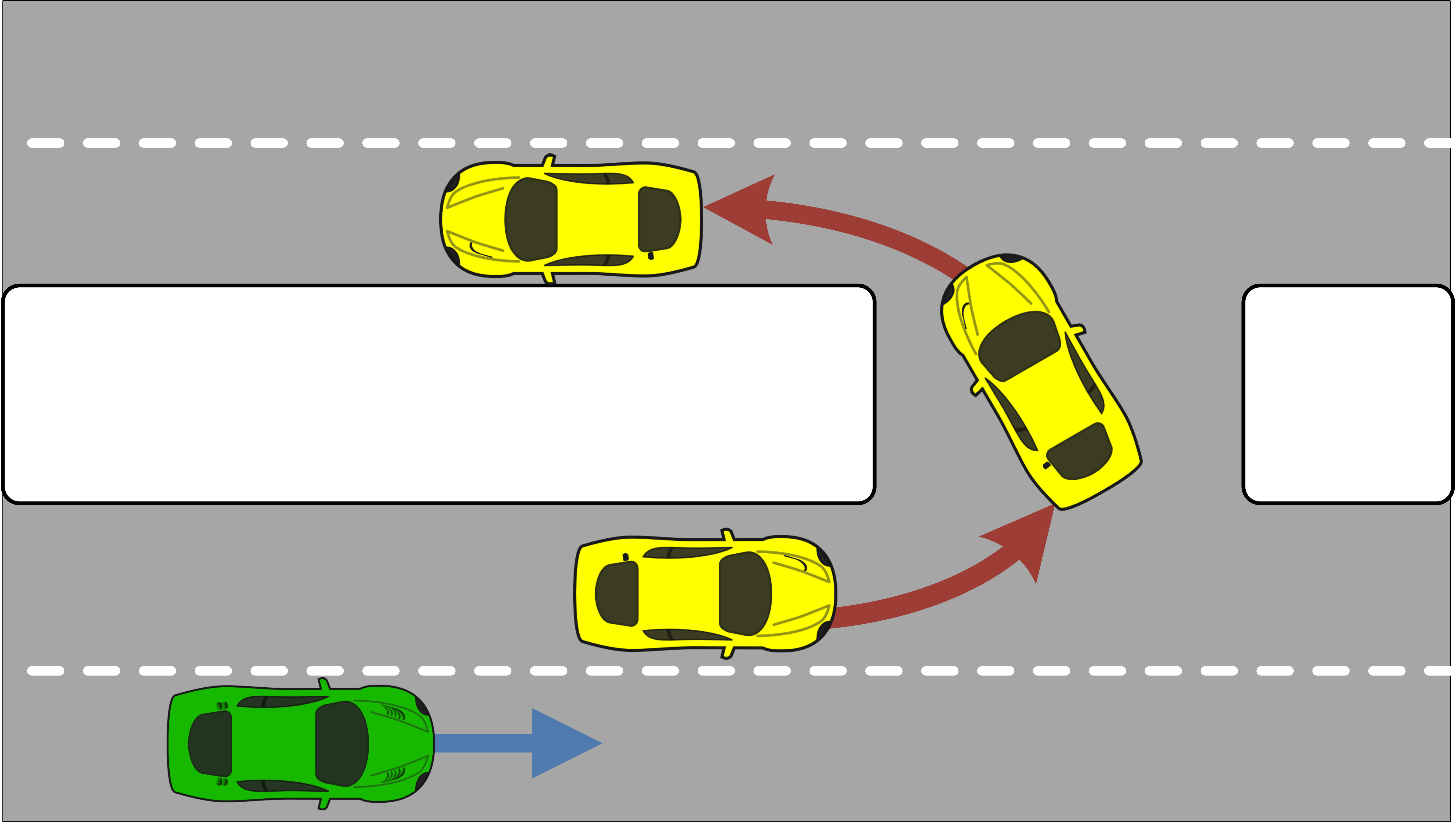}       \label{fig:uturn_map}
    }
    \hspace{2pt}
    \subfigure[U-turns cause a change around $180^\circ$ in the estimated orientation.]{
      \includegraphics[width=0.4\linewidth]{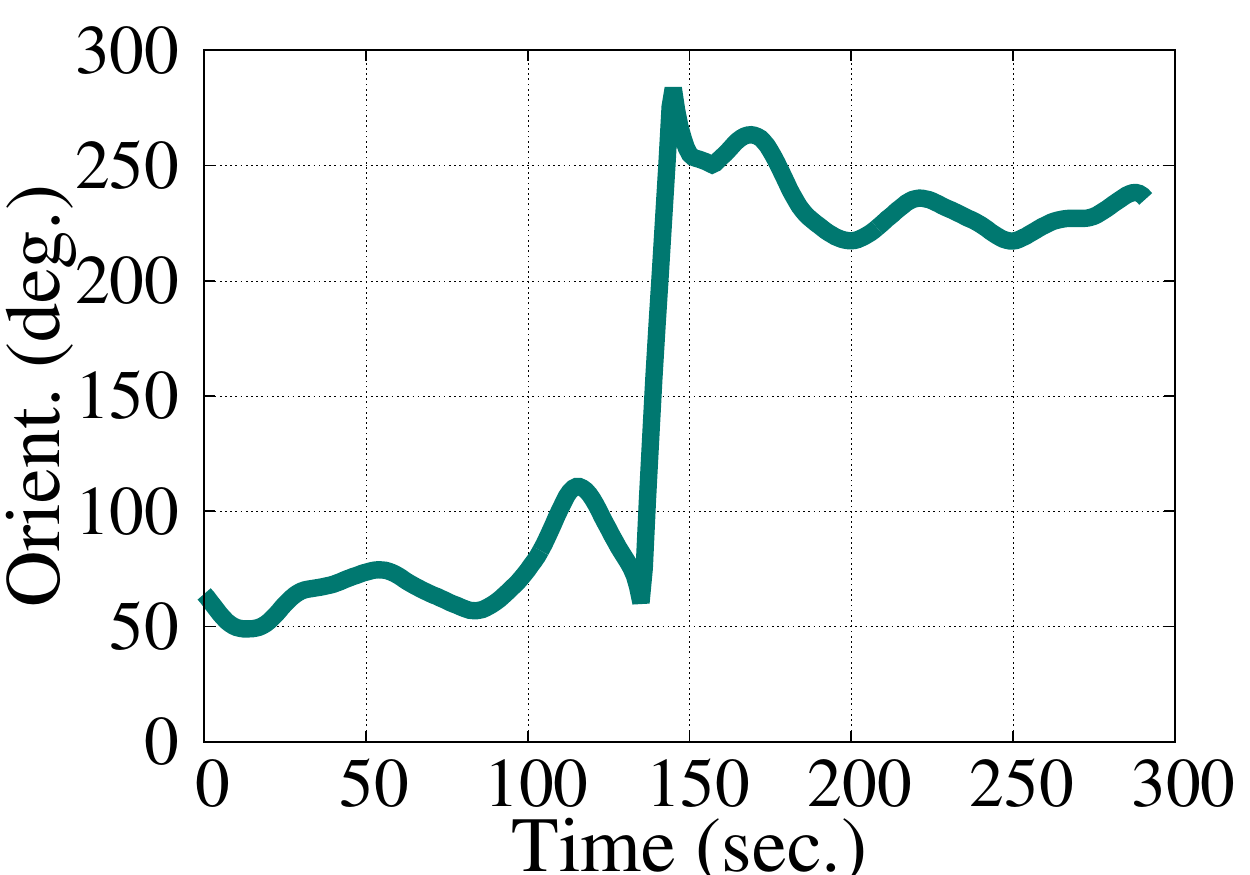}
  \label{fig:uturn_orient}
    }
\caption{The phones' orientation sensor can detect a u-turn, which in turn gives a better idea about the car current lane.}
\label{fig:uturn_ex}
\end{figure}

Similarly, when making a u-turn, the car is most probably at the left-most lane before and after the u-turn. Therefore, noting that the car's direction changes by around $\pm 180^\circ$ when making a turn, which can be captured using the cellphone's orientation sensor (Figure~\ref{fig:uturn_ex}) using the \emph{Lane Anchor Detection} module, this ``u-turn anchor'' hint is used by \sys{} to reduce the ambiguity of the car's current lane.

\sys{} differentiates between two types of lane anchors: \textbf{Bootstrap anchors} and \textbf{Organic anchors}. \emph{Bootstrap anchors} have a clear \emph{pre-known} lane distribution across the road. For example, stopping a car occurs in the right-most lane; a u-turn is initiated in the left-most lane, and a right-turn happens with high probability in the right-most lane. On the other hand, \emph{organic anchors} have unique signatures across the different lanes but their lane distribution \emph{cannot be pre-known} and need to be learned. For example, a pothole can be detected by the phone sensors as we show in Section~\ref{sec:system}. However, we do not know a priori in which lane this pothole is located. \sys{} uses an unsupervised crowd-sourced approach to capture these anchors and identify their lanes distribution. 
The details of operation of this module are discussed in Section~\ref{sec:event_detect}.
\subsection{Probabilistic Lane Estimation Module}
To achieve robust and accurate lane estimates based on the noisy inertial sensors measurements, the ambiguous car locations, human driving anomalies, and fuzzy lane anchor locations; \sys{} uses a probabilistic estimation technique. Specifically, our lane estimation technique is based on Markov Localization~\cite{russell1995artificial,burgard1999markov}, which is known in the robotics domain for addressing the problem of state estimation from noisy sensor data. Instead of maintaining a single hypothesis about the robot location, Markov localization uses a probabilistic
framework to maintain a probability density over the set of possible locations. Such a density can have an arbitrary form representing various position beliefs, including multimodal distributions. Markov localization can deal with ambiguous situations and it can re-localize the robot position in the case of localization
failures. The basic assumption in Markov localization is that the current state, i.e. the current robot location, captures the entire movement history (Markov assumption). That is, the current position is the only state in the environment which systematically affects the sensors readings.

Accordingly, \sys{} uses Markov localization to maintain a probability distribution over all possible lanes. This probabilistic representation allows it to weigh the different hypotheses and reach a more accurate lane estimate in a mathematically principled way. \sys{} does not make any assumption regarding the starting lane position of the car. This is modeled as a uniform distribution across all lanes. Then, as the car moves on the road, any  cues for the car motion (i.e. lane changes) or detected lane anchors (e.g. a pothole) are used to update this lane belief distribution (Figure~\ref{fig:markov_ex}).  For example, assuming a car is moving on a four-lane road and it made three right lane changes, each  time a lane change is detected the car's lane position distribution is updated. After the third lane change, the car is at the right-most lane with high probability. Similarly, if we know that the road has a pothole at the second lane around the current car location and the car encounters it, then most probably it is at the second lane.

The details of operation of this module are discussed in Section~\ref{sec:prob_est}.
\subsection{Organic Lane Anchors Updates Module}
This module is responsible for estimating the location and lane distribution of organic anchors such as curves and potholes. It uses a crowd-sensing approach, where the information about the detected lane anchors from different system users is collected and processed to estimate the anchor location and lane distribution based on the reporting cars' lane distributions.

The details of operation of this module are discussed in Section~\ref{sec:organic_update}.
\section{The \system{} System}\label{sec:system}
In this section, we provide the details of the \sys{} novel probabilistic lane estimation, event detection module, our unsupervised lane-anchors detection algorithm, and practical considerations. We start by the mathematical notations.
\subsection{Notations}
\begin{itemize}
\item Let $\ell_t$ denote the actual car's lane position at time $t$ and $L_t$ denote the corresponding discrete random variable. $\ell_t$ can take values from $1$ to $n$; where $n$ is the number of road lanes.
\item The belief about the car's lane position at time $t$ is $Bel(L_t)$. $Bel(L_t)$ is the probability mass function representing the probability distribution over the road's lanes. \item Let $e_t$ denotes the event detected at time $t$. The system can detect two types of events: motion events $m_t$ (i.e. lane changes to the right or left) and lane anchor detection event $a_t$ (e.g. a pothole or a u-turn).
\item Let $s_t$ denote the car's lane position estimate at time $t$ using our probabilistic algorithm.

\end{itemize}
\subsection{Probabilistic Lane Estimation}\label{sec:prob_est}
Our lane estimation module aims to estimate the car's lane using the detected events (motion events and anchors detection events). Since the lane belief ($Bel(L_t)$) changes only when the phone sensors detect an event, then at time $T$, the car's lane belief will be based on all detected events till that time ($e= e_0,e_1,...,e_T$).  This corresponds to the posterior distribution over the road's lane conditioned on all the detected events, that is
\vspace*{-10pt}
\begin{equation}
Bel(L_t)= P(L_t=\ell|e) = P(L_T=\ell|e_0,...e_T)
\label{eq:output1}
\vspace*{-4pt}
\end{equation}
When computing  $P(L_t=\ell|e)$, we have two cases based on the two event types (motion event and anchor detection event).
\\
\\
\\
\textbf{Case 1: The event is a motion event, i.e. lane change ($e_T=m_T$):}
\\
A car moving over the road will use the same lane till it makes a right ($m^r$) or a left ($m^l$) lane change.
 Therefore, when the phone sensors detect a lane change (as in Section~\ref{subsec:lane_change_det}), the lane belief in
Eq.~\ref{eq:output1} can be factorized to:
\begin{equation}
\resizebox{\columnwidth}{!}{
$Bel(L_T)= P(L_T=\ell|e) = \sum_{i=1}^{n} P(L_T = \ell|e,L_{T-1}=\ell_i)P(L_{T-1}=\ell_i|e)$
\label{eq:total_pr}
}
\end{equation}
This is based on the theorem of Total Probability~\cite{ang2004probability}, where the probability that the car is at a certain lane $\ell$ as a result of a lane change event is mapped to the summation of the possibility of being at any previous lane position multiplied by the transition probability of moving to $\ell$ from this previous lane.
\\
From the Markovian assumption, we can further simplify the term $P(L_T = \ell|e,L_{T-1}=\ell_i)$ to:
\begin{equation}\resizebox{\columnwidth}{!}{$\displaystyle \begin{array}{ll}
P(L_T=\ell|e,L_{T-1}=\ell_i) & = P(L_T = \ell|e_0,...,e_{T-1},m_T,L_{T-1}=\ell_i)\\
 &= P(L_T=\ell|m_T,L_{T-1}=\ell_i)\\
 \end{array}
$}\end{equation}
Similarly, $m_T$ should not affect the lane position at $T-1$ in the term $P(L_{T-1}=\ell_i|e)$. Hence Eq.~\ref{eq:total_pr} can be written as:
\begin{equation}
\resizebox{\columnwidth}{!}{$\displaystyle \begin{array}{ll}
Bel(L_T=\ell)= P(L_t=\ell|e) & \\
= \sum_{i=1}^{n} P(L_T=\ell|m_T,L_{T-1}=\ell_i)P(L_{T-1}=\ell_i|e_0,...,e_{T-1})&
 \end{array}
 $}
\end{equation}
Which can be written in a recursive form as:
\begin{equation}
\resizebox{\columnwidth}{!}{
$Bel(L_T=\ell) = \sum_{i=1}^{n} P(L_T=\ell|m_T, L_{T-1}=\ell_i)Bel(L_{T-1}=\ell_i)$
}
\end{equation}
The probability $P(L_T=\ell|m_T, L_{T-1}=\ell_i)$ represents the motion model and it should capture the uncertainty in our sensors measurements. We model this uncertainty using the
confusion matrix between left-lane-change ($m^l$), right-lane change ($m^r$) and no-lane change ($m^0$) (Section~\ref{subsec:lane_change_det}). For example, if the current detected motion event is a left lane change, then $P(L_T=\ell|m_T=m^l,L_{T-1}=\ell_i)$ is calculated as:
\vspace*{-13pt}
\begin{equation}\resizebox{\columnwidth}{!}{$\displaystyle \begin{array}{cr}
P(L_T=\ell|m_T=m^l,L_{T-1}=\ell_i) = &
\left\{
\begin{array}{ll}
      P(m^l|m^l) & \ell = \ell_{i}+1\\
      P(m^l|m^r) & \ell = \ell_{i}-1\\
      P(m^l|m^0) & \ell = \ell_{i}\\
       0 & o.w.\\
\end{array}
\right.
\end{array}
$}\label{eq:motion_r}
\end{equation}
\vspace*{5pt}
\\
\textbf{Case 2: The event is observing an anchor ($e_T=a_T$) (perception model):}
\\
The second type of events we have in our system is passing by one of the lane-anchors, i.e. $e_T=a_T$.
In this case, Eq.~\ref{eq:output1} can be factorized using Bayes' rule to:
\begin{equation}
\resizebox{\columnwidth}{!}{
$P(L_t=\ell|e) = \frac{P(a_T|e_0,...,e_{T-1},L_T=\ell)P(L_T=\ell|e_0,...,e_{T-1})}{P(a_T|e_0,...,e_{T-1})}$
}
\end{equation}
That can be simplified based on our Markov assumption to:
\begin{equation}
P(L_t=\ell|e) = \frac{P(a_T|L_T=\ell)P(L_T=\ell|e_0,...,e_{T-1})}{P(a_T|e_0,...,e_{T-1})}
\label{eq:bayesian}
\end{equation}
Noting that the denominator of the last equation does not depend on $L_T$, we can replace it by a constant $(\alpha_T)$ (i.e. a normalizing factor). Therefore, Eq.~\ref{eq:bayesian} becomes:
\begin{equation}
P(L_t=\ell|e) = \alpha_TP(a_T|L_T=\ell)P(L_T=\ell|e_0,...,e_{T-1})
\end{equation}
Again, this can be put in a recursive form as:
\begin{equation}
Bel(L_t=\ell) = (\alpha_T)P(a_T|L_T=\ell)Bel(L_{T-1}=\ell)
\end{equation}
The term $P(a_T|L_T=\ell)$ represents the perception model, which is the likelihood that the lane-anchor's signature $a_T$ would be observed if the user was actually in lane ($\ell$). Two factors affect this model: whether there is actually an anchor of the detected type near the car current location and the anchor lane distribution. Therefore, we model this probability as a weighted Gaussian distribution as follows:
\begin{equation}
P(a_T|L_T=\ell) = P(\ell|a_T) \frac{1}{\sqrt{2\pi}\sigma}e^{-0.5\left(\frac{|\ell-\ell_{a_T}|}{\sigma}\right)^2}
\end{equation}
Where $|\ell-\ell_{a_T}|$ is the distance between the $a_T$ anchor's lane position $\ell_{a_T}$ and the car current lane position $\ell$ and $\sigma$ is the uncertainty in the sensors measurements and the anchor's lane position. We estimate $\sigma$ as the median absolute deviation (MAD) which is a robust estimator of $\sigma$ \cite{newson2009hidden}.
\begin{equation}
\sigma = 1.4826 \times \text{median}_T(|\ell - \ell_a|)
\end{equation}
Finally, the current car lane ($s_T$) is estimated as the lane with the maximum belief.

Our lane estimation algorithm is summarized in Algorithm~\ref{alg:markov_laneest}.
\begin{algorithm}[!t]\small
\caption{Lane Estimation Algorithm}
\begin{algorithmic}[1]
\For{\textbf{each} $\ell$}
\State $Bel(L_0 = \ell) \gets 1/n$\Comment{Initialize the lane belief} \EndFor \While{$true$}
\If{an anchor is detected}
\State $\alpha_T \gets 0$
\For{\textbf{each} $\ell$} \Comment{Perception model}
\State $\hat{Bel}(L_T = \ell) \gets P(a_T|\ell)Bel(L_{T-1} = \ell)$
\State $\alpha_T \gets \alpha_T + \hat{Bel}(L_T = \ell)$
\EndFor
\For{\textbf{each} $\ell$} \Comment{Normalize the lane belief}
\State $Bel(L_T = \ell) \gets \alpha_T^{-1} \hat{Bel}(L_{T-1} = \ell)$
\EndFor
\EndIf
\If{lane change detected}
\For{\textbf{each} $\ell$} \Comment{Motion model}
\State $Bel(L_T=\ell) \gets \sum_{i=1}^{n} P(\ell|\ell_i,m_T)\hat{Bel}(L_{T-1}=\ell_i)$
\EndFor
\EndIf
\State $s_T \gets \underset{\ell}{\arg\max}\text{ } Bel(L_T=\ell)$ \Comment{Estimate current lane}
\EndWhile
\end{algorithmic}
\label{alg:markov_laneest}
\end{algorithm}
\subsection{Events Detection}\label{sec:event_detect}
In this section, we describe how \sys{} detects the motion events (i.e. lane change) and the anchor detection events.
\subsubsection{Lane Change Detection}\label{subsec:lane_change_det}
\begin{figure}[!t]
\centering
\includegraphics[width=0.63\linewidth]{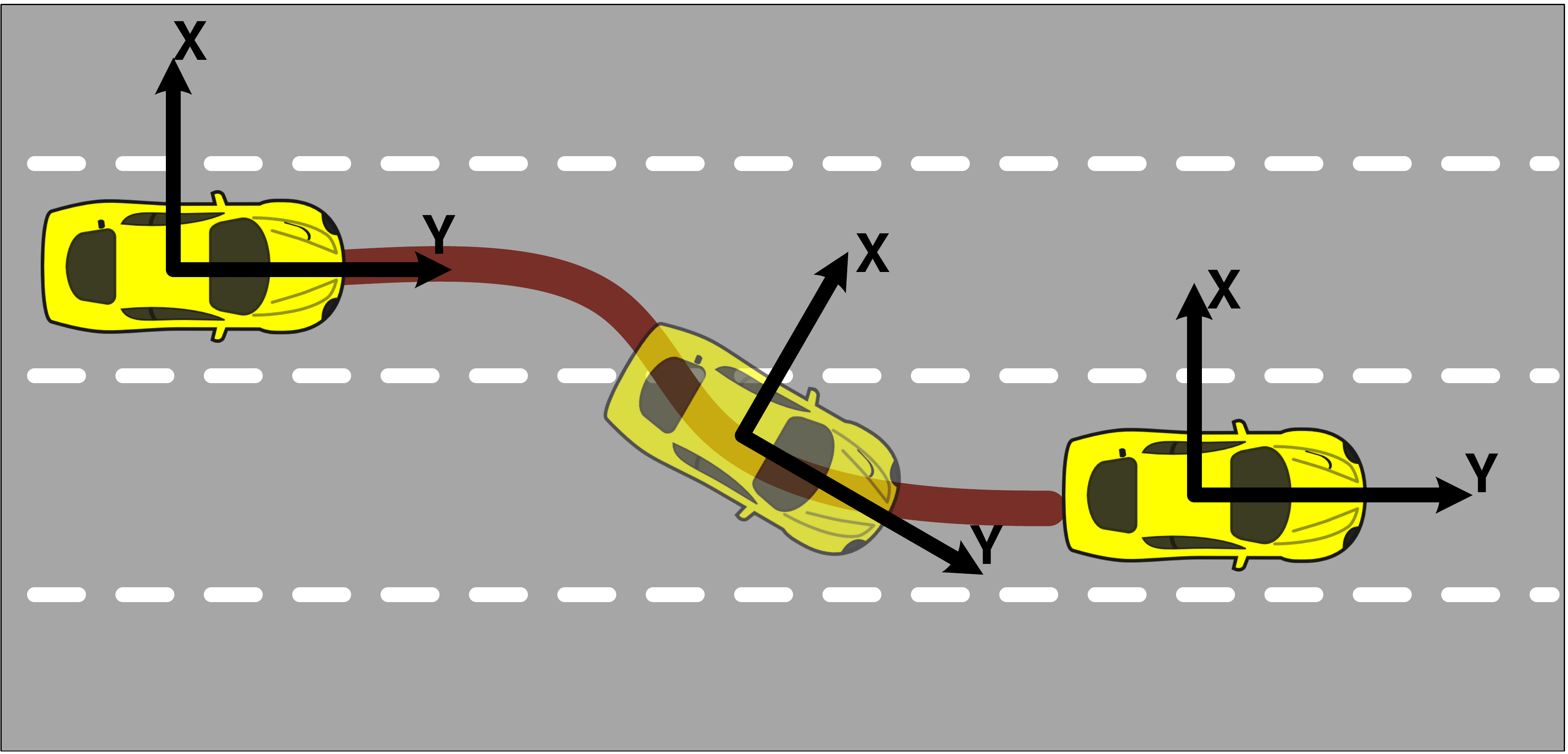}
\caption{A car doing a lane change will have to make a small rotation around the z-axis, leading to a change in its x-axis acceleration.}
\label{fig:lane_change_ex}
\end{figure}

\begin{figure}[!t]
    \centering
 \subfigure[Left lane change]{
      \includegraphics[width=0.38\linewidth]{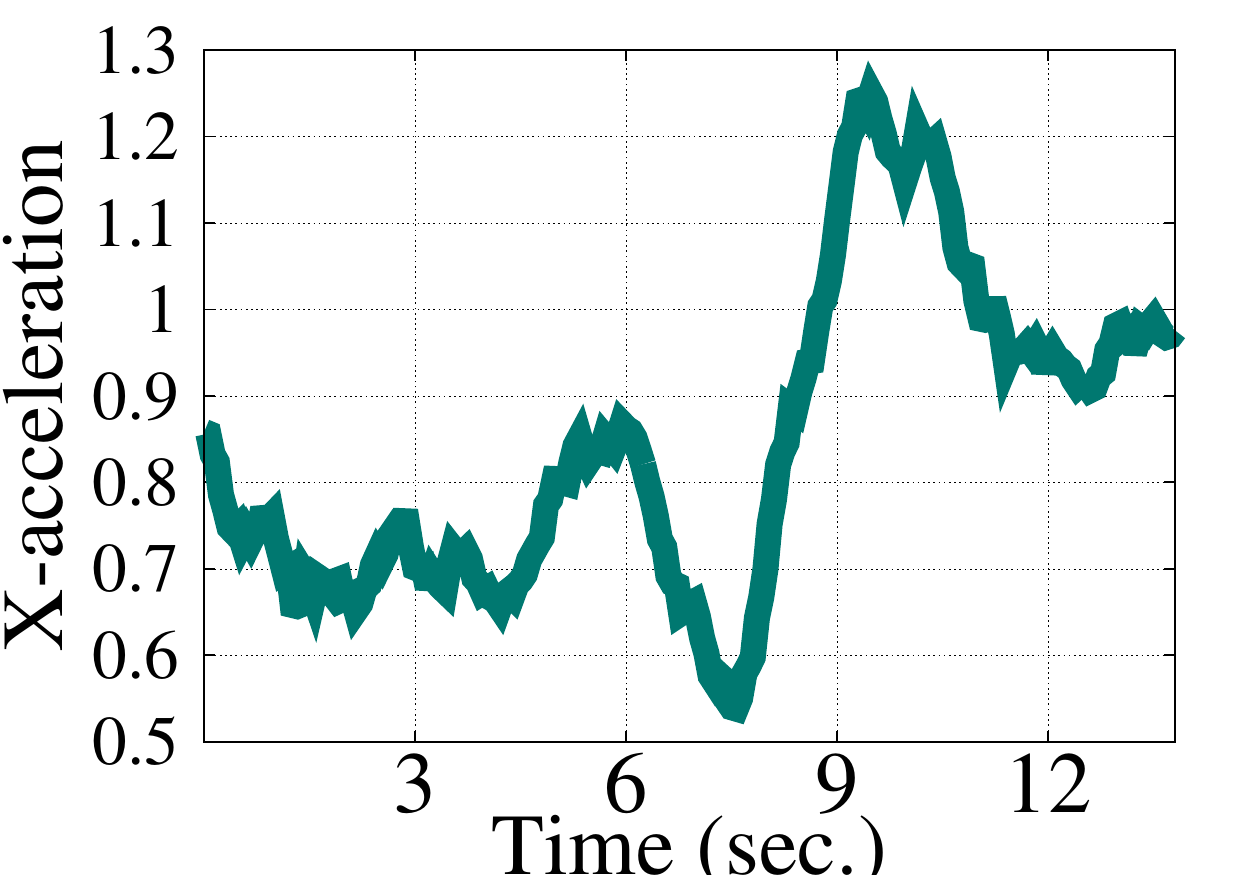}
      \label{fig:xacc_l}
    }
    \hspace{2pt}
    \subfigure[Right lane change]{
      \includegraphics[width=0.38\linewidth]{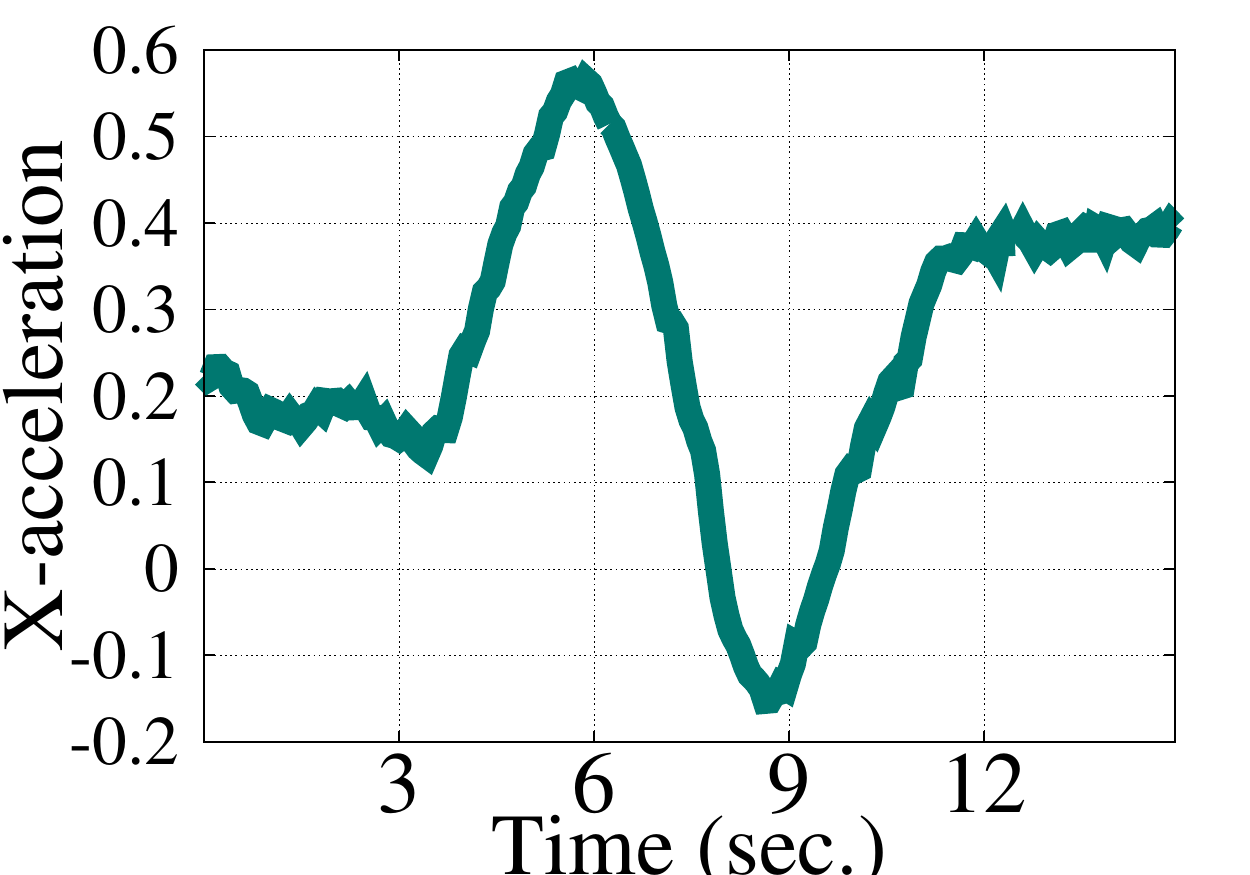}
  \label{fig:xacc_r}
    }
\caption{A left lane change causes a specific pattern on the x-acceleration. The pattern is reversed for right lane change.}
\label{fig:lane_change}
\end{figure}

Drivers typically change their lanes while driving for several reasons including: a) the current lane is ending/merging b) the driver plans to make a turn at an upcoming intersection, or c) the driver wants to move to a faster/slower moving lane. A number of techniques in literature proposed using the phone inertial sensors to detect the car lane change event \cite{fazeen2012safe,li2012marvel}. The idea is that for the car to change its lane, it experiences a change in its direction (Figure~\ref{fig:lane_change_ex}), which causes a rotation around the z-axis of the accelerometer (for the oriented phone) and affects mainly the x-acceleration\cite{serway2013physics}. Assuming that the car is making a left-lane change (Figure~\ref{fig:xacc_l}), then the x-acceleration reading first decreases to a low value and then increases back to a higher value. The pattern is reversed for the right lane change as shown in Figure~\ref{fig:xacc_r}. To capture this pattern, we use a slightly modified version of the approach proposed by~\cite{li2012marvel}, where we detect the maximum and minimum peaks within a window. If the difference between the two peaks exceeds a threshold and they are close (within 4 seconds~\cite{li2012marvel}), we detect a lane change event. We detect whether it is a left or a right lane change from the peaks order (Figure~\ref{fig:lane_change}). Based on the traces from our experiments, we found that a window size of $7$ sec. and a threshold of $0.8 m/s^2$ gave good results (Section~\ref{sec:eval}).
\begin{figure*}[!t]
\centering
\includegraphics[width=0.85\linewidth]{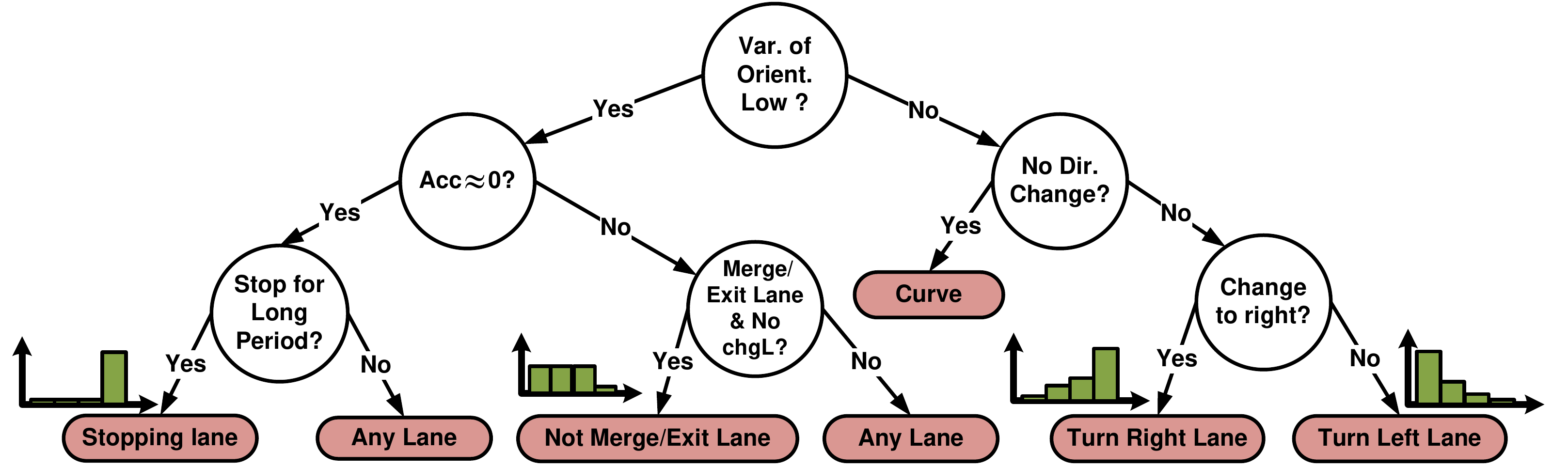}
\caption{The decision tree used to identify \textbf{the bootstrap lane anchors}.}
\label{fig:laneknown}
\end{figure*}
\subsubsection{Bootstrap Lane Anchors}
\sys{} defines bootstrap anchors as anchors that have unique sensors signature and a priori known lane distribution. These anchors include turns, merging and exit lanes, and stopping lanes. For the rest of this subsection, we will give details about the anchors detection and lane distribution for each of them. Figure~\ref{fig:laneknown} shows the decision tree used to identify the bootstrap lane-anchors.

\noindent\textbf{\textit{Turns:}}
Turns and u-turns force the car to change its direction by around $90^\circ$ and $180^\circ$ respectively, which results in a big variance in the car's orientation along with a change in its final orientation when it ends. This can be captured using the phone's orientation sensor as shown in Figure~\ref{fig:uturn_orient}. To further differentiate between right and left turns, the difference between the starting and ending direction can be computed or the x-acceleration can be used as it results in patterns similar to the lane-change event (Figure~\ref{fig:lane_change}).

Since the driver should make a turn only from the closest lane, the lane distribution for turn anchors is a skewed distribution according to the turn type. This distribution can be used initially and updated dynamically based on the crowd-sensed data as discussed in the next section.

\noindent\textbf{\textit{Merging and Exit Lanes:}}
A merging lane is used to merge traffic between two roads. Similarly, an exit lane is used to exit a road, e.g. a highway, to another. Usually these lanes have a special extra lane to the main lanes on the road. The location of these lane anchors can be extracted from the digital map and passing by them can be detected based on the car's map matched location.  These lanes are usually the last lanes to the right or left. Therefore, if a car uses an exit or merge lane, its lane distribution will be skewed. Note also that not taking an exit or merge lane indicates that the car is not located in these special lanes. This negative information can be associated with the complement distribution of this type of anchors.
\noindent\textbf{\textit{Stopping Lanes:}}
A car may only park in the right-most lane of a driving road. However, traffic signals and road congestion can make a car stop at any lane. To differentiate between parking and the other cases, we use a simple time filter, where parking is detected only if the car stops for more than 3 minutes.

A parking anchor distribution clusters mainly on the right-most lane only and have small weights for the other lanes.

     \begin{figure}[!t]
\centering
\includegraphics[width=0.5\linewidth]{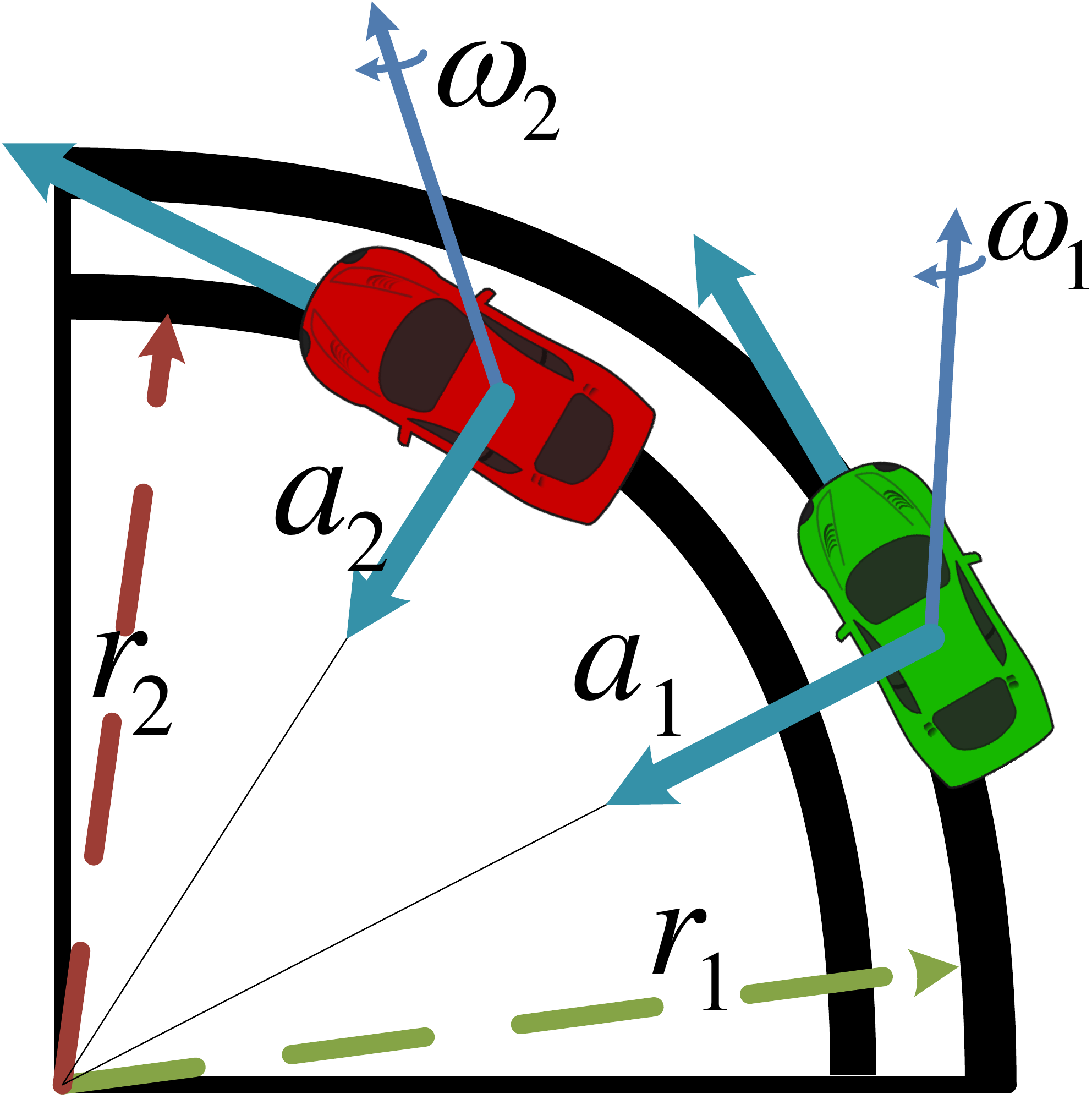}
\caption{While moving over a curved road with radius $r_i$, the magnitude of the centripetal acceleration ($a_i$) is related to the angular velocity ($\omega_i$).}
\label{fig:curve_rule}
\end{figure}
\begin{figure}[!t]
    \centering
      \includegraphics[width=\figscale\linewidth]{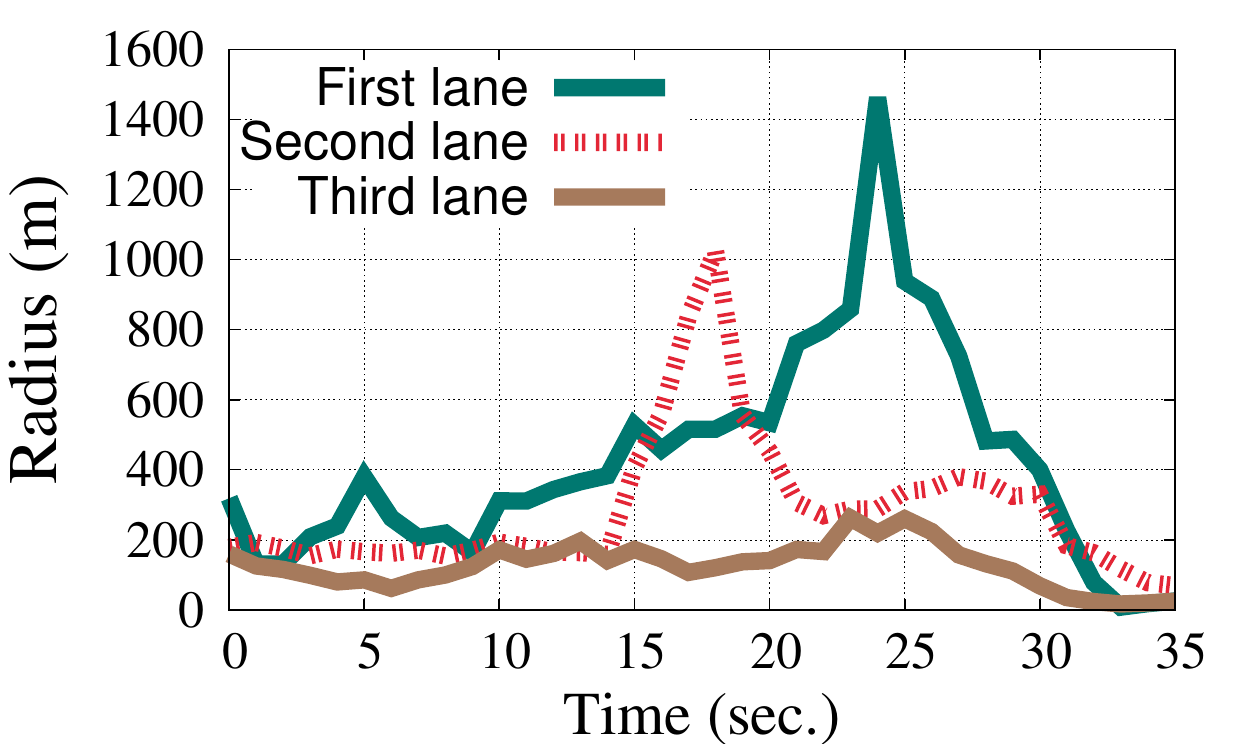}
      \caption{Estimated radius for a car moving at the different lanes of the same curve. The outer lane (1st lane) has the largest radius and the radius decreases for inner-lanes. }
\label{fig:curve_rad}
\end{figure}

\begin{figure}[!t]
\centering
\includegraphics[width=\figscale\linewidth]{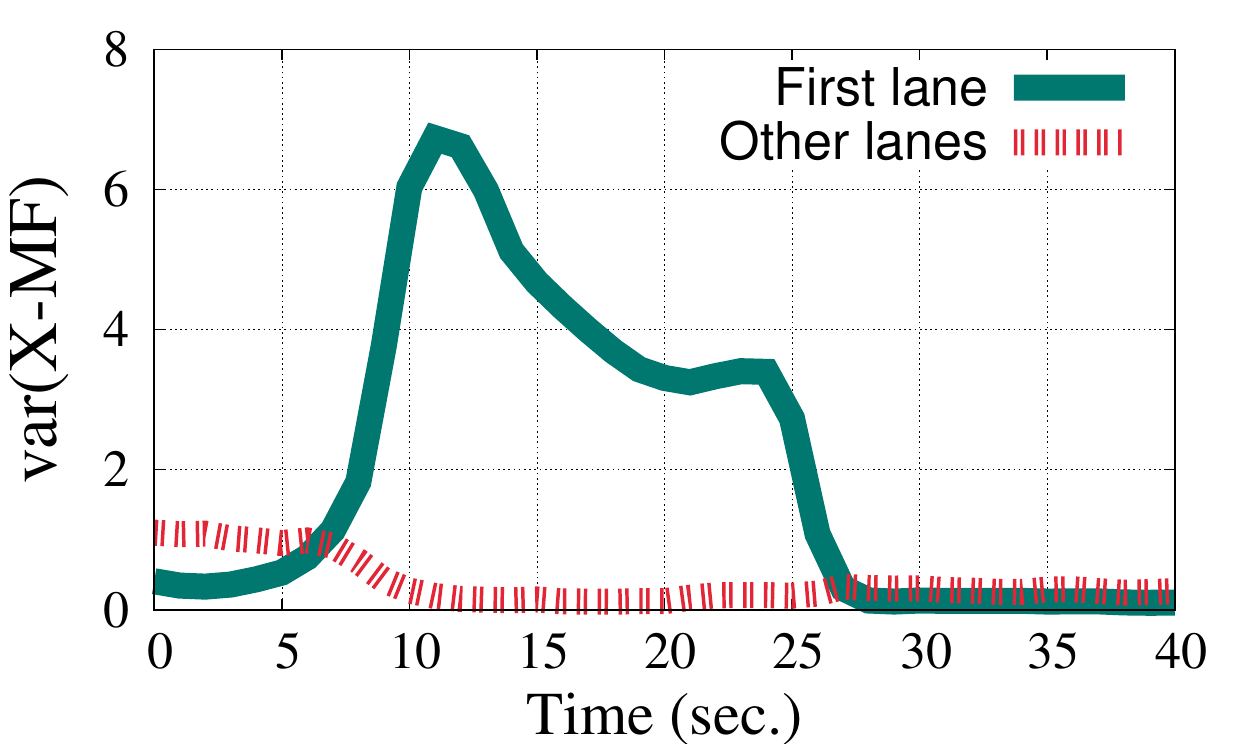}
\caption{\small As the car goes inside/outside the tunnel, it experiences a higher variance in the x-magnetic field which decreases as we move away from the lane close to the infrastructure.
}
\label{fig:tunnel_lanes_ex}
\end{figure}
\subsubsection{Organic Lane Anchors}\label{sec:curve_lane}
\sys{} also defines organic anchors which have unique sensors characteristics across the different lanes. However, their lane distribution and road position cannot be predetermined without war-driving. These anchors include curves, tunnels, and potholes. For the rest of this subsection, we will give details about these anchors unique characteristics and how they differ across the different lanes. We leave the details of learning their characteristic in an ``organic'' way to the next subsection.

\noindent\textbf{\textit{Curves:}}
When a car drives over a curved-road with radius $r$, the direction of its tangential velocity vector ($\nu$) changes as it rotates over the curve. The rate of the direction change is the centripetal acceleration ($a$), which always points inwards along the radius vector of the circular motion. Without this acceleration, the car would move in a straight line, according to Newton's laws of motion. Based on the circular motion laws~\cite{serway2013physics}, the magnitude of the centripetal acceleration ($a$) is related to the angular velocity ($\omega$) as (Figure~\ref{fig:curve_rule}):
\begin{equation}
a = \omega^2 r
\label{eq:cent_acc}
\end{equation}
Which can be arranged as:
\begin{eqnarray}
r = \frac{a}{\omega^2}\label{eq:omega_r}
\vspace*{-6pt}
\end{eqnarray}
This equation provides a methods for estimating the radius of the lane the car is moving in based on the inertial sensors (angular velocity and centripetal acceleration).
Figure~\ref{fig:curve_rad} shows an example of the radius estimated for road curves at the different lanes. We can see a clear distinction between them.  
\noindent\textbf{\textit{Tunnels:}}
Going inside a tunnel causes a drop in the cellular signals for all the heard cell-towers \cite{aly2013dejavu}. This drop can be used to detect the tunnel, but not the specific lane inside the tunnel as it is sensed in all lanes. Studying the effect of moving inside large tunnels with a number of lanes, we noticed a large variance in the ambient magnetic field in the x-direction (perpendicular to the car direction of motion) while the car is going inside the tunnel and going out of the tunnel. This can be explained by the metal and infrastructure (e.g. electricity lines) that exist on the side of the tunnel structure. This high variance decreases as you move away from the tunnel's side where the infrastructure is installed (Figure~\ref{fig:tunnel_lanes_ex}). This is expected as magnetic interference is known to have an effect on smart-phone's magnetometer within small distances only \cite{chung2011indoor}. 
\noindent\textbf{\textit{Potholes and other anomalies:}}
Anomalies in the road surface such as potholes span only part of the road compared to traffic calming device (e.g. bumps and cat's eyes), which spans the whole road. We identify such anomalies
using thresholding on the variance of the z-gravity acceleration as in \cite{mednis2011real}. However, this leads to an ambiguity with other traffic calming devices. To resolve this ambiguity, we further use our unsupervised learning approach described in the next section. Typically, a traffic calming device such as a bump will have a uniform distribution over all lanes compared to a pothole that has a narrow distribution. 
\begin{figure}[!t]
\centering
\includegraphics[width=\figscale\linewidth]{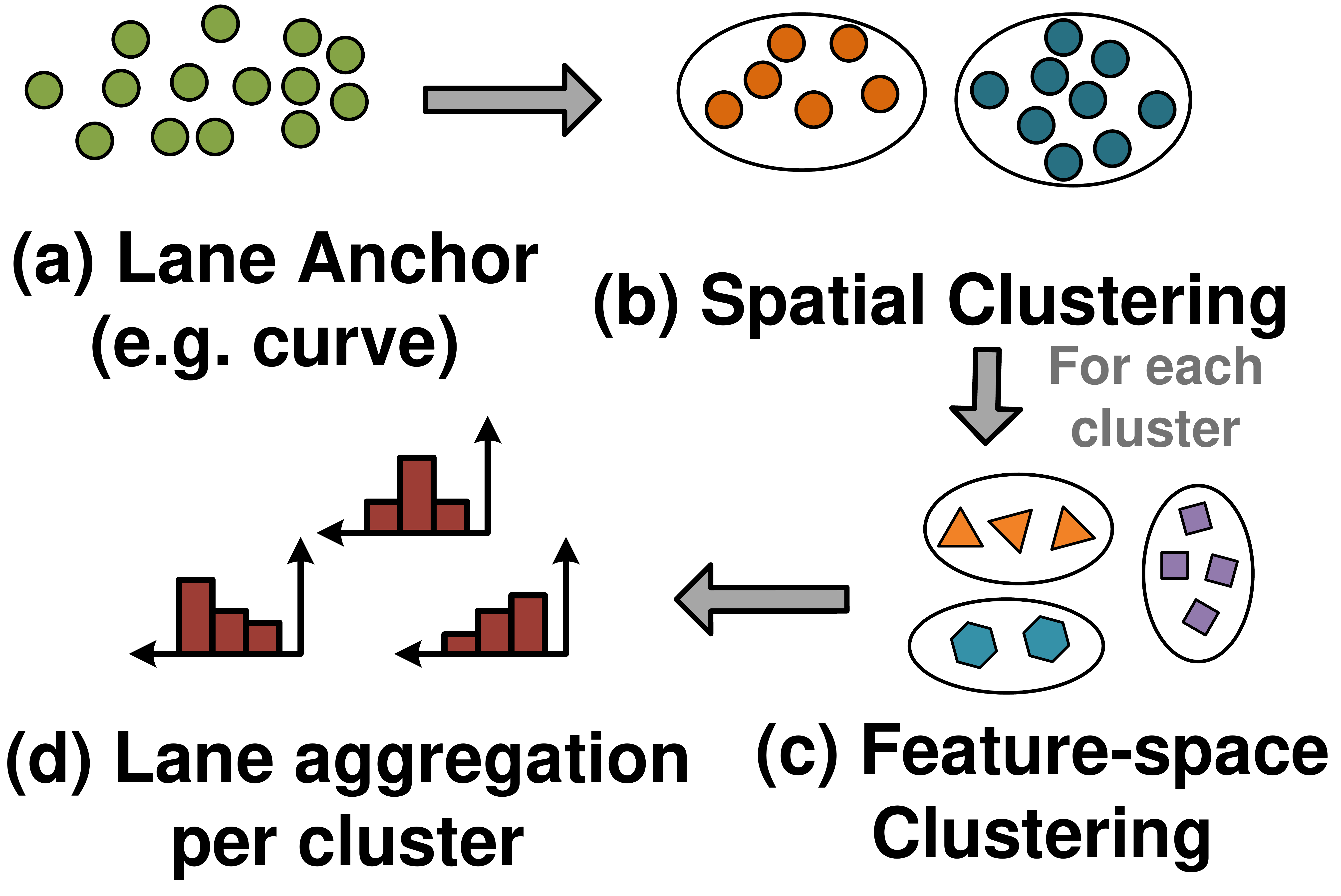}
\caption{The three step unsupervised crowd-sourcing  approach used by \sys{} to learn the road location and lane distribution of the organic lane anchors.}
\label{fig:ulearn}
\end{figure}
  \begin{table*}[!t]
  \centering
  \resizebox{0.9\linewidth}{!}{
  \begin{tabular}{||c||c||c|c|c||c|c|c||c||c|c||}\hline\hline
  \multirow{2}{*}{Testbed} &  \multirow{2}{*}{ Distance Covered}& \multicolumn{3}{c||}{Speed (Km/h)}& \multicolumn{3}{c||}{Number of Lanes}& \multirow{2}{*}{Number of Roads} &   \multicolumn{2}{c||}{Number of Lane Anchors} \\\cline{3-8}\cline{10-11}
  & & Min. & Avg. & Max. & Min & Avg. & Max.& &Bootstrap & Organic\\\hline\hline
  Alexandria, EG & 200& 0&40.3&70.0 & 3 & 4.6 &5 &22&359&312\\\hline  Makkah, KSA & 60&0&88.3&106.2 &  3&3.1&4&8&45&48\\\hline\hline  \end{tabular}  }  \caption{Summary of the different testbeds used.}
      \label{tab:testbed}
  \end{table*}

\subsection{Organic Lane Anchors Automatic Detection}\label{sec:organic_update}
Organic anchors have known sensors signature but
their exact location in the road and their probability distribution across
the lanes cannot be predetermined unless a calibration phase across the area of interest is employed. Typically, this imposes an arduous data collection at the different lanes for the entire area.  To reduce this overhead, we propose an unsupervised crowd-sourcing approach for identifying these lane-anchors profile. Specifically, for each identified road-anchor (e.g. a curve lane), we aim to determine its road location as well as its lane span distribution. 
We use a three step process to determine the lane anchor profile (Figure~\ref{fig:ulearn}). Without loss of generality, we use the curve lane anchor as an example. \emph{First}, we apply \emph{spatial} clustering on all samples collected from all users that are detected as curves. This separates the different curves over the area of interest. The road location of the lane anchor is taken as the centroid of all points within this cluster. \emph{Second}, for each resulting cluster (representing one specific curve), we do a second level clustering of its points based on the lane-discriminating features (the radius in this case as explained in Section~\ref{sec:curve_lane}) to separate the curve lane-anchors\footnote{Note that for a given curve, each radius corresponds to a different lane anchor.}. This helps in determining the lane position for a given lane anchor. \emph{Third}, the curve lane anchor probability distribution ($P(a|\ell)$) is constructed from all the reported car lane beliefs for the points within this last resulting feature-based clustering step. In particular, we take the average of lane beliefs of the different points inside the cluster as:
\begin{equation}
P(a|\ell) = \frac{1}{c}\sum_{i=1}^{c} P(L=\ell|u_i)
\label{eq:aggr}
\end{equation}
Where $c$ is the number of points inside the cluster and $P(L=\ell|u_i)$ denotes the probability that car $u_i$ was at lane $\ell$ when it passed by this lane anchor. Similarly, other lane anchors are identified using the same process. The lane-discriminating features for the different lane-anchors are explained in detail in Section~\ref{sec:curve_lane}.

Finally, we note that we choose a density-based clustering algorithm (DBSCAN~\cite{ester1996density}) for the two level clustering algorithms as the number of clusters is not required to be known in advance, the detected clusters can have arbitrary shapes, and outliers can be detected. 
\subsection{Practical Considerations}~\label{sec:discuss}
\subsubsection{Adherence to traffic rules}
Some drivers may violate traffic rules occasionally. For example, a driver can make a right turn from the second right-most lane. \sys{}, since it uses a probabilistic framework, can incorporate this possibility in its lane distribution by flattening the lane distributions of the different anchors (e.g. by increasing the distribution variance). Also, many lane anchors, e.g. curves and potholes, are traffic rules-independent which can correct and update inaccuracies due to these violations; leading to a high overall lane estimation accuracy as we quantify in Section~\ref{sec:eval}.
\subsubsection{Number of lanes}
\sys{} assumes that the used digital map includes information about the number of lanes for the different roads. Such information is available in a number of the current digital maps  and can also be automatically inferred using crowd-sourcing approaches, e.g. \cite{chen2010probabilistic}.
\begin{figure}[!t]
\centering
\includegraphics[height=3cm,width=\figscale\linewidth]{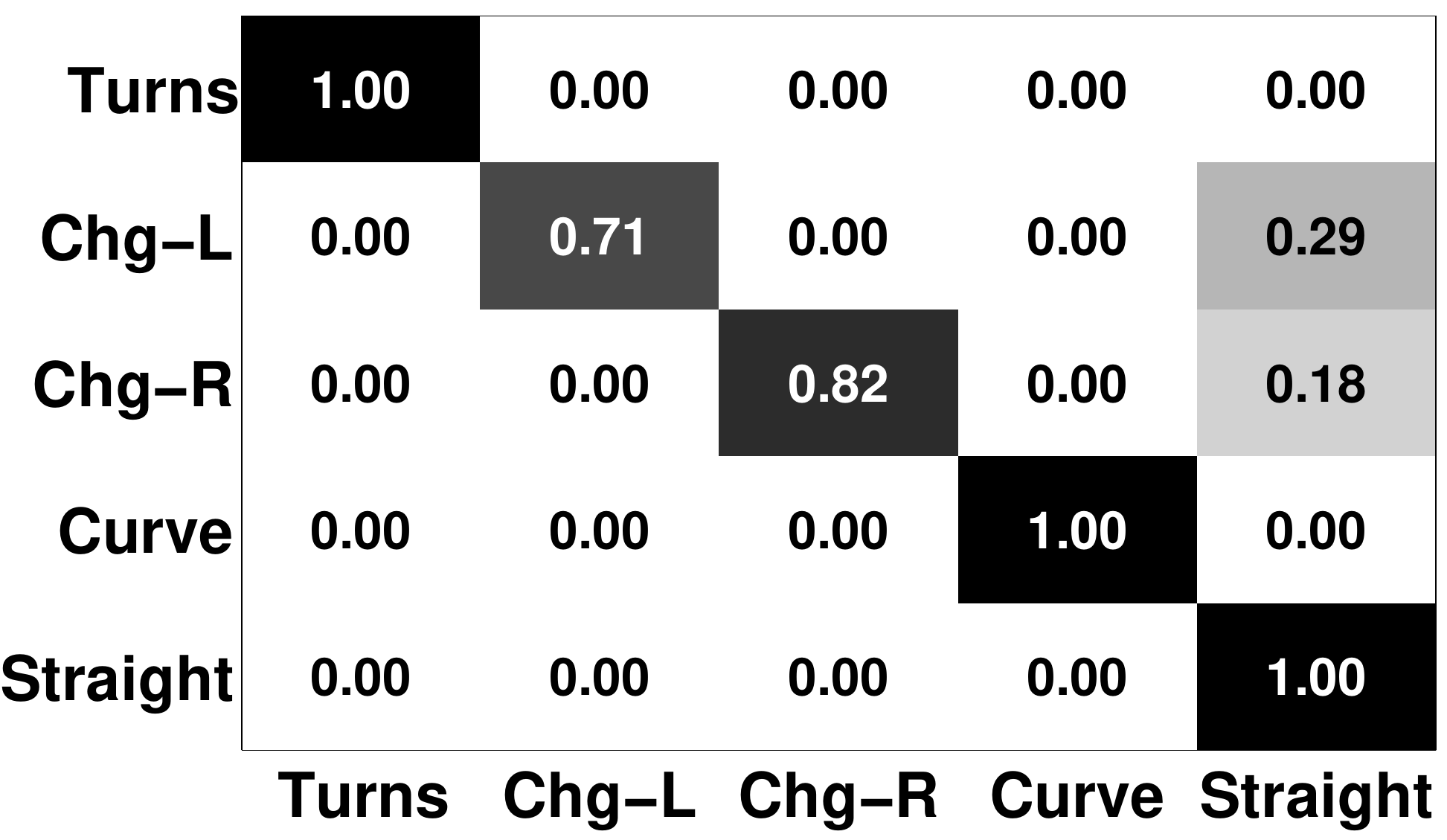}
\caption{Confusion matrix for the lane change events and the related anchors (curve).   Note that Chg-L/R denotes left/right lane change.}
\label{fig:dyn_conf}
\end{figure}
 \begin{figure}[!t]
\centering
\includegraphics[width=\figscale\linewidth]{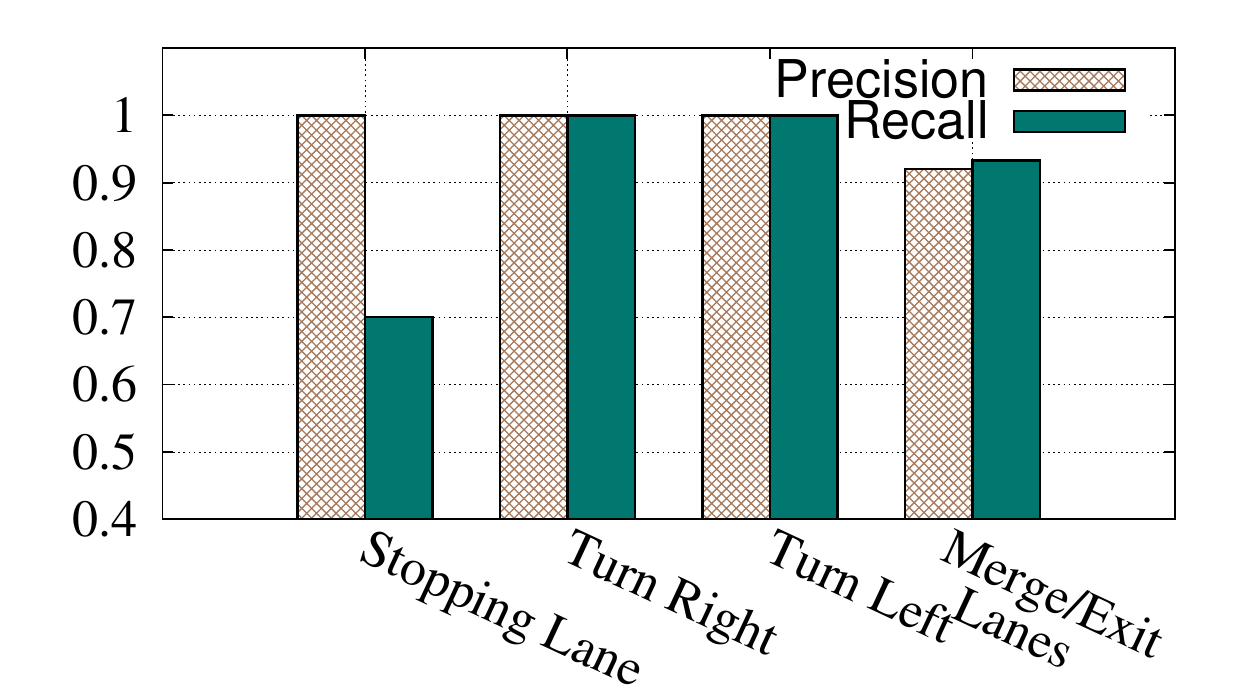}
\caption{\sys{} can identify the different \textbf{bootstrap lane anchors} with an average precision and recall of \textbf{0.98} and \textbf{0.91} respectively.}
\label{fig:anchor_det_boot}
\end{figure}

\section{Evaluation}\label{sec:eval}

We implemented \sys{} on different android devices including LG Nexus 4, HTC M8, Samsung Galaxy Note, Samsung Galaxy S4, and Samsung Galaxy Nexus. We evaluated the system in \emph{the city of Alexandria, Egypt and the city of Makkah, Saudi Arabia} using 13 drivers with a combined drive traces length of \emph{200 km in Alexandria and 60 km in Makkah}. On average, we had about 4.2 lanes per road in our traces. Table~\ref{tab:testbed} summarizes the testbeds parameters.

To define our lane position ground truth, we developed a simple application to facilitate marking lane positions across the trips and the number of lanes in a particular road. The application has a map with a pointer for the user current position. To set the lane position, the user clicks on the pointer and choose the number of her current lane. To reduce the error due to manual labeling of the ground truth, the application detects lane changes and prompts the user to confirm the lane change and the new lane position. Without loss of generality, we use GPS as the localization technique through this section.

For the rest of this section, we start by evaluating the accuracy of identifying the different events and anchors. Then we show the overall lane estimation accuracy. Finally, we show the energy-consumption overhead of adding \sys{} to different localization systems to estimate the car's lane.

\subsection{Motion Events Detection Accuracy}
Figure~\ref{fig:dyn_conf} shows the confusion matrix for detecting the motion events (i.e. lane change) and the related anchors (i.e. turns and curves). The matrix shows that we can detect lane-changes, turns, and curves with high accuracy. This in turn enables high accuracy in lane estimation as we quantify later.
\begin{figure}[!t]
\centering
\includegraphics[width=\figscale\linewidth]{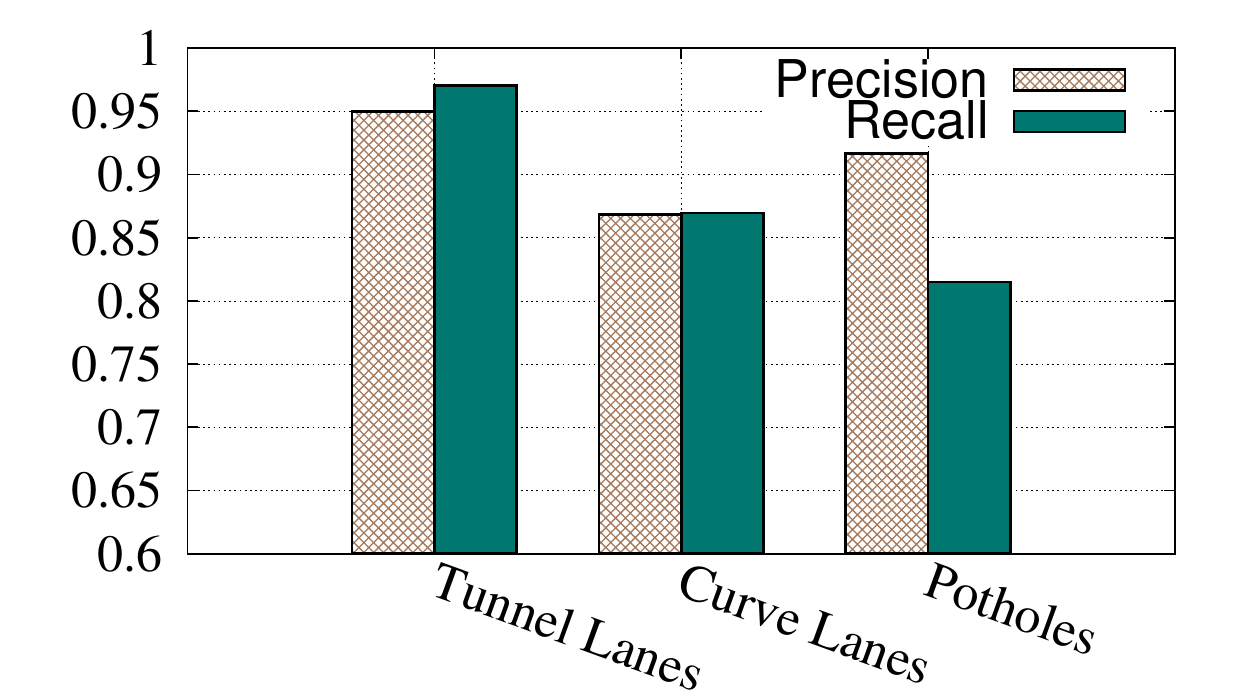}
\caption{\sys{} can identify the different \textbf{organic lane anchors} with an average precision and recall of \textbf{0.91} and \textbf{0.88} respectively.}
\label{fig:anchor_det_org}
\end{figure}
\begin{figure}[!t]
\centering
\includegraphics[height=3.5cm,width=\figscale\linewidth]{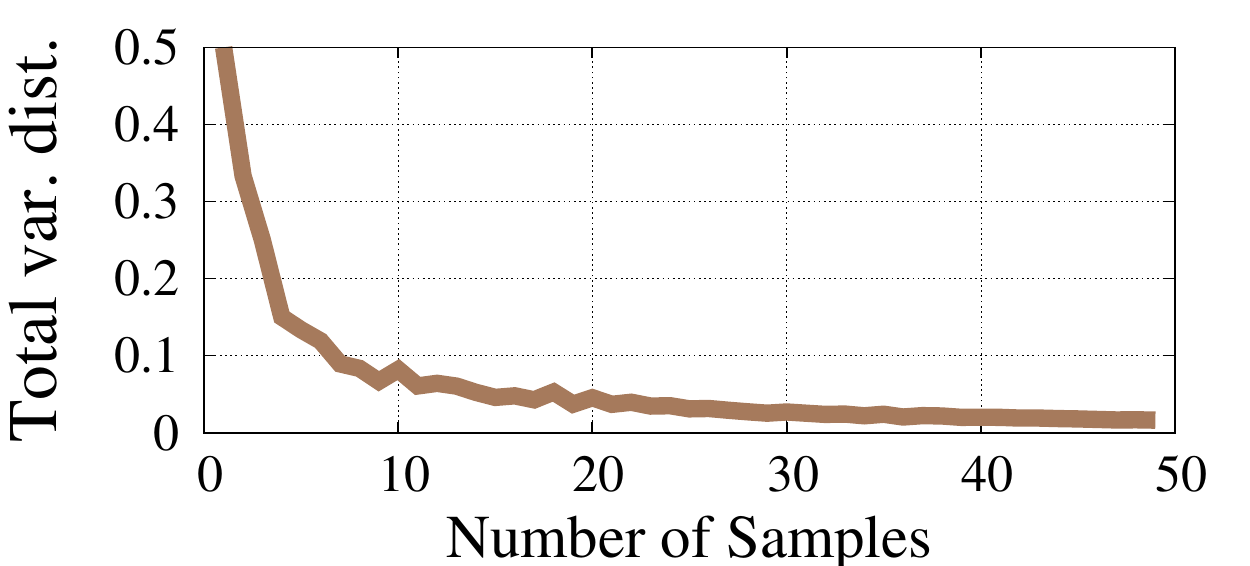}
\caption{Effect of number of points within a cluster of traces on the lane-anchors identification accuracy.}
\label{fig:bump_idf}
\end{figure}
\subsection{Lane Anchors Detection Accuracy}
\subsubsection{Bootstrap Lane Anchors Detection}
Figure~\ref{fig:anchor_det_boot} provides the precision and recall for the different bootstrap lane anchors. The figure shows that we can identify the different lane anchors accurately with an average precision and recall of 0.98 and 0.91 respectively.
\subsubsection{Organic Lane Anchors Detection}
Figure~\ref{fig:anchor_det_org} provides the precision and recall for the different organic lane anchors. Note that curves here reflect the accuracy of detecting the correct lane within the curve as opposed to separating the curve from other events in the confusion matrix. The figure shows that that we can identify the different organic lane anchors accurately with an average precision and recall of 0.91 and 0.88 respectively.

Figure~\ref{fig:bump_idf} shows the effect of the number of points within a cluster of traces on the
organic lane-anchors identification accuracy reflected by a zero total variation distance~\cite{levin2009markov} between successive distributions. The figure shows that our two-stage clustering algorithm converges to a stable lane anchor distribution using as few as 20 points per organic anchor. This number is even amortized over the different cars that pass by this specific anchor. 

\subsection{Lane Estimation Accuracy}
\begin{figure}[!t]
\centering
\includegraphics[width=\figscale\linewidth]{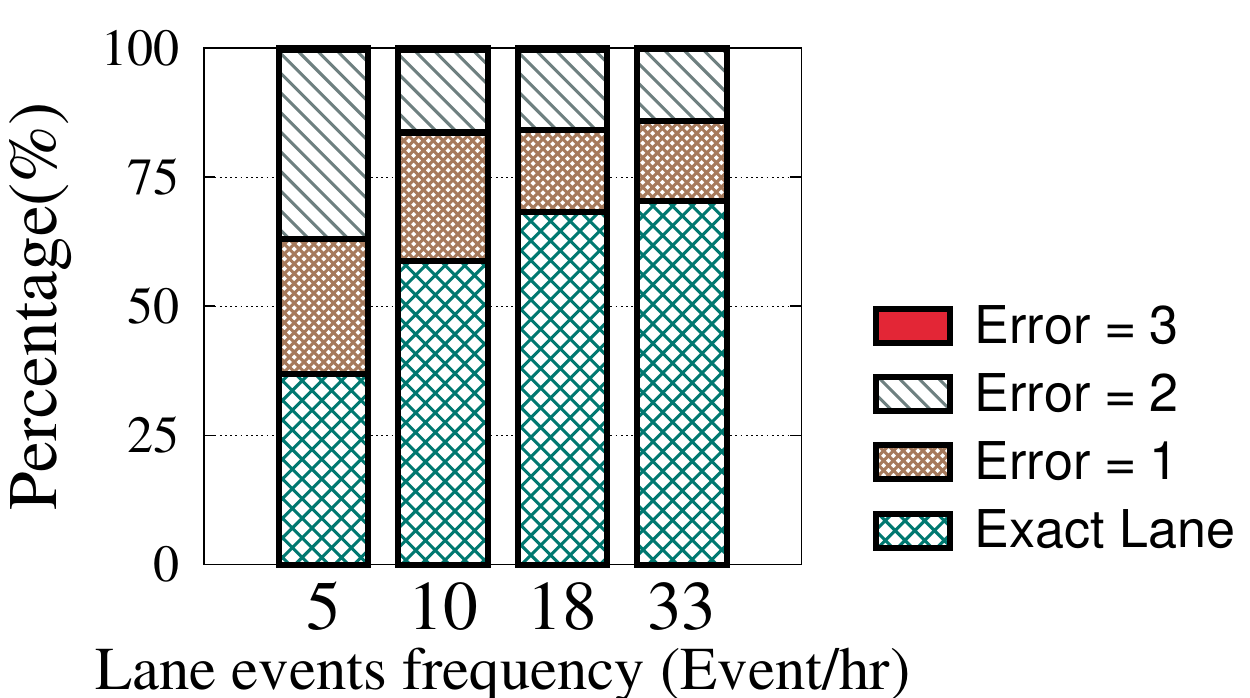}
\caption{Effect of events frequency on lane estimation accuracy.}
\label{fig:cdf_density}
\end{figure}
\subsubsection{Effect of lane event frequency on accuracy}
While we expect a user to detect a large number of lane-events (e.g. lane-changes, curves, turns, etc.), typically, we cannot predict how many lane-events the user will encounter during a trip. For this, we study the effect of reducing the frequency of detected lane anchors on accuracy by sub-sampling the actual detected events (Figure~\ref{fig:cdf_density}). The figure shows that even with a low rate of 10 lane-events detections per hour, \sys{} can still identify the car's exact lane more than 60\% of the time. 
\begin{figure}[!t]
\centering
\includegraphics[width=\figscale\linewidth]{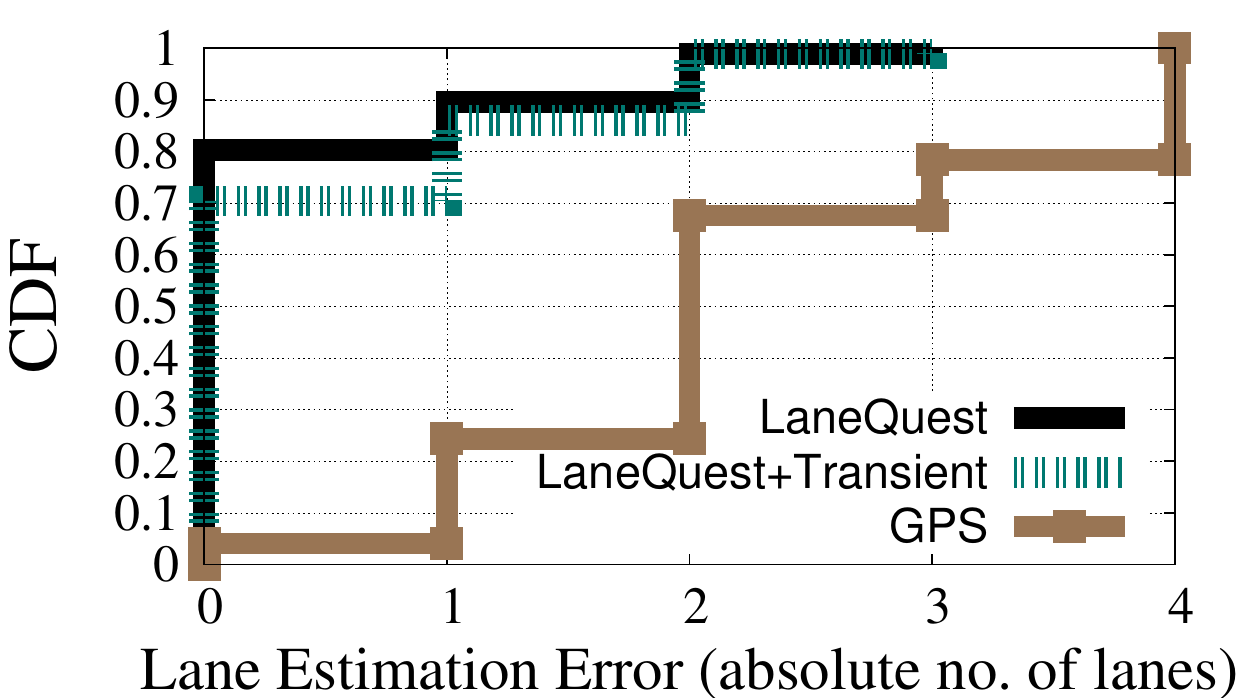}
\caption{CDF of lane estimation error when using \sys{} compared to GPS.}
\label{fig:error_cdf}
\end{figure}
\subsubsection{Steady state system accuracy}
Figure~\ref{fig:error_cdf} shows the CDF of the lane estimation error for \sys{} (with and without the transient period) compared to GPS. For GPS, we take the lane estimate as the closest lane to the reported GPS location. Due to the GPS inaccuracy, GPS biases its lane estimate to the rightmost or leftmost lane, leading to a large error in lane estimation. On the other hand, \sys{} can identify the car's exact lane more than 70\% of the time. This increases to 89\% to within one lane error. Moreover, since we start \sys{} from an unknown lane position, it incurs the highest errors at the beginning. After removing this transient stage, \sys{} performance increases by $15\%$.  On average, the transient stage was in order of few minutes. However, using \sys{} from the beginning of the trip shortens it  to less than a minute (due to detecting the stopping-lane anchor, which has a clear and peaked distribution).
\begin{figure}[!t]
\centering
\includegraphics[width=\figscale\linewidth]{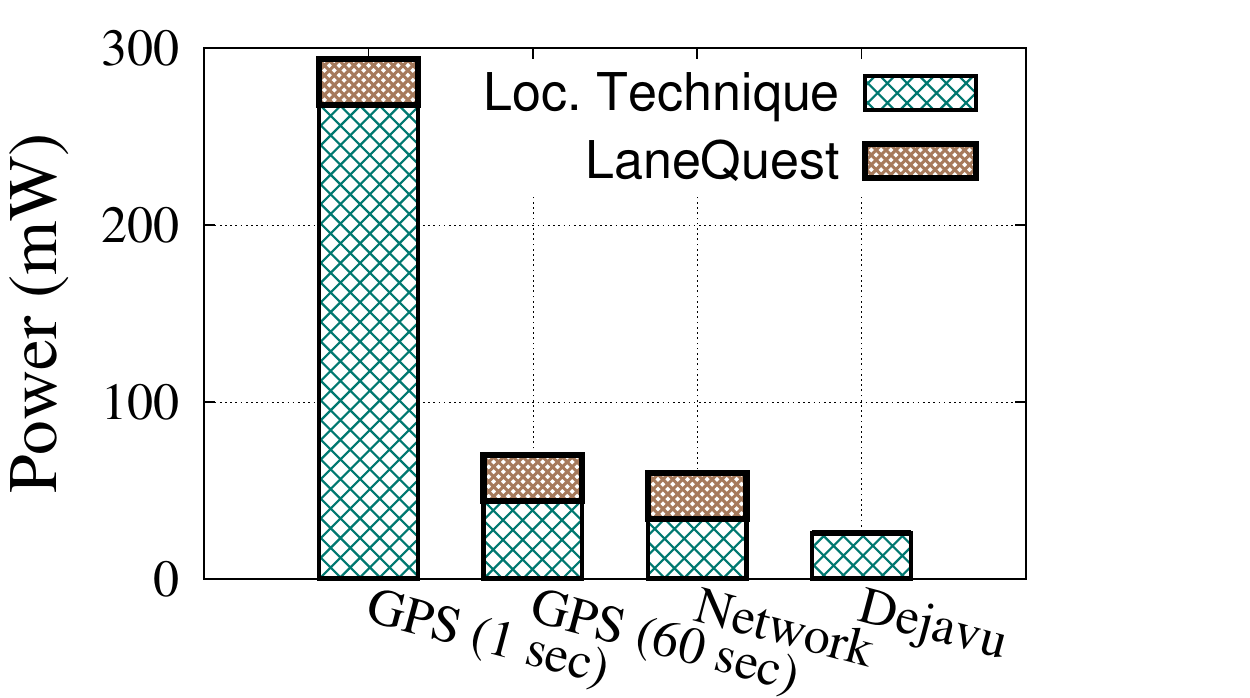}
\caption{Power consumption for the different systems when integrated with \sys{}.}
\label{fig:power}
\end{figure}
\subsection{Energy Overhead}
Figure~\ref{fig:power} shows the energy overhead when integrating \sys{} with other localization systems. The power consumption were calculated using the PowerTutor profiler~\cite{zhang2010accurate} and the android APIs using the HTC Nexus One cellphone. Even though we implemented \sys{} on GPS only, we compare its energy consumption with other localization systems (mainly WiFi-based localization and the Dejavu system~\cite{aly2013dejavu}) based on estimating their energy consumption from the sensors they use. The figure shows that \sys{} has a small negligible energy footprint. In addition, when combined with systems that use the inertial sensors for localization, e.g. Dejavu~\cite{aly2013dejavu}, it consumes zero extra energy. This highlights its suitability for use with the energy-constrained mobile devices.
\section{Related Work}\label{sec:relwork}\sys{} provides an accurate lane estimation using inertial sensors available in commodity smart-phones. It leverages a set of lane-dependent driving patterns along with different road anchors to identify the car's lane. Through this section, we discuss the different lane-level localization techniques and previous techniques that use inertial sensors for sensing different road parts or driving behavior.
\subsection{Lane Determination}
Current state-of-the-art localization techniques can only provide location estimates with an average accuracy around 10m~\cite{aly2013dejavu}, which is not suitable for lane-level localization. To overcome this, researchers proposed different techniques~\cite{toledo2010lane,dao2007markov,selloum2009lane} that are based on using an external accurate GPS device (L1 GPS or DGPS) for localization and combining it with other sensors to provide more accurate lane-level location estimates. For example, in~\cite{toledo2010lane} authors fuse a high-accuracy GNSS receiver, an odometer, and a gyroscope, along with an enhanced digital map (that describes the road geometry using a particle filter-based algorithm) to position the user based on a combined GNSS-dead-reckoning approach and map matches her to her lane. Similarly, in~\cite{tao2013lane} authors fused an L1-GPS device, camera and an enhanced digital map that stores lane markings information using a dynamical Kalman filter with map matching to get an accurate user position.

All these techniques require special hardware, sometimes with ubiquitous deployment, in order to function; which limits their applicability on a large scale.

Computer vision techniques have been proposed to detect the car lane. For example, in  \cite{ren2010lane} authors use the iPhone camera to detect the lane markings. Using cameras for lane detection, however, is highly susceptible to errors due to various factors such as lighting condition (e.g. night time, sun glare, headlight glare, shadows from nearby buildings, etc), bad weather conditions (e.g. snow, rain), and other environmental noise (e.g., faded lane marks, surrounding objects like buildings, parked cars, etc). Moreover, both camera and GPS have high energy requirements for the limited phone battery.

\sys{}, on the other hand, provides an accurate lane estimate using only energy-efficient inertial sensors, eliminating the need for any special devices to be installed on the car or an expensive pre-calibrated enhanced digital map. In addition, it works well in many weather and environment conditions.

\subsection{Road Sensing and Driving Behavior Detection}
Inertial sensors have been used in literature for detecting: driving behavior~\cite{fazeen2012safe,wang2013sensing,singh2013using}, map semantics~\cite{aly_map14,sheikh2014demonstrating}, and road problems~\cite{pothole,mednis2011real}. For example, in~\cite{fazeen2012safe,singh2013using} authors used inertial sensors to detect the driving quality of the car's driver. They identified driving patterns events like lane-changing and acceleration/deceleration and rated the driver according to the frequency and suddenness of these events. Similarly, in~\cite{wang2013sensing}, authors used inertial sensors and an external accelerometer to sense the car dynamics when turning to detect the driver's phone usage. \sys{} leverages similar events for estimating the car's lane with extensions to separate close events, such as making a turn or moving on a curve, for more robust and accurate lane localization.

In~\cite{aly_map14}, authors inferred various map-semantics to enrich digital maps such as tunnels, roundabouts, and bridges among others using cellphone's inertial sensors and cellular-information.
In~\cite{mednis2011real}, authors used the cellphone accelerometer to detect the potholes without separating them from normal traffic calming devices, e.g. bumps. The Pothole Patrol~\cite{pothole} system, on the other hand, uses a 3-axis accelerometer and GPS to detect potholes along the road and apply a series of filters on the acceleration to separate between potholes and others like bumps and expansion joints. However, it uses \emph{external} accelerometer which has higher sampling rate and lower noise compared to chips available on typical cellphones in the market.

\sys{}, on the other hand, uses the \emph{cheap noisy inertial sensors} in standard cellphones to detect driving patterns like changing lanes or road semantics like tunnels. More importantly, it identifies more fine-grained anchors \textbf{\emph{at the lane-level}}, e.g. instead of just detecting one anchor for the tunnel, we detect different anchors for the lanes inside the tunnel. In addition, it uses an \textbf{unsupervised} crowd-sourcing approach to learn the signatures of these lane-level anchors.

\section{Conclusion}\label{sec:conclude}We presented the \sys{} system for providing an accurate estimate of the car's lane position. \sys{} depends only on energy-efficient inertial sensors available on commodity off-the-shelf smart phones. It detects the car's lane without any prior assumption on its starting lane position based on a novel probabilistic framework that fuses knowledge of the car's dynamics with lane-anchors. These anchors are learned through an organic crowd-sourcing approach.

Implementation of \sys{} on a number of android devices using typical driving traces at different cities shows that it can detect the different lane-level anchors with an average recall and precision of more than 90\%. This helps it detect the correct lane accurately with more than 80\% of the time, increasing to 89\% of the time to within one lane error. This comes with a low energy profile, allowing it to be implemented on the energy-constrained mobile devices.

Currently, we are extending the system in multiple directions including experimenting with other motion and perception models, extracting more lane-level anchors, using other phone sensors, among others.
\section*{Acknowledgment}
This work was supported in part by the KACST National
Science and Technology Plan under grant \#11-INF2062-10,
and the KACST GIS Technology Innovation Center at Umm
Al-Qura University under grant \#GISTIC-14-02.

\bibliographystyle{IEEEtran}
\bibliography{lanes}
\end{document}